\documentclass[twocolumn]{emulateapj}
\usepackage[latin2]{inputenc}

\usepackage{amssymb}
\usepackage{amsmath}
\usepackage{txfonts}
\usepackage{amsfonts}

\def\hMpc{{\ }h^{-1}{\,}{\rm Mpc}}
\def\ihMpc{{\ }h{\,} {\rm Mpc}^{-1}}
\def\hMpcnosp{h^{-1}{\,}{\rm Mpc}}

\def\hGpcnosp{h^{-1}{\,}{\rm Gpc}}

\def\xisdss{\xi_{\rm SDSS}}
\newcommand{\avg}[1]{\left\langle{#1}\right\rangle}
\newcommand{\avgnol}[1]{\langle{#1}\rangle}
\def\xh {\hat\xi}
\def\tsky {\theta_{\rm sky}}
\def\pbuf{P_{\rm Buffon}}
\def\nore{N_{\rm orient}^{\rm eff}}
\def\nor{N_{\rm orient}}
\def\noc{N_{\rm occupied}}
\def\lcdm{$\Lambda$CDM}
\def\camb{{\scshape camb}}

\begin{document}

\title{Redshift-space Enhancement of Line-of-Sight Baryon Acoustic
  Oscillations in the SDSS Main-Galaxy Sample}

\author{H.\ J.\ Tian\altaffilmark{1,2,3}, Mark C.\ Neyrinck\altaffilmark{2}, 
Tam\'as Budav\'ari\altaffilmark{2}, Alexander S.\ Szalay\altaffilmark{2}}

\altaffiltext{1}{The Institute of Astrophysics, Huazhong Normal
University, Wuhan 430079, China} \altaffiltext{2}{Department of
Physics and Astronomy, Johns Hopkins University, Baltimore, MD 21218}
\altaffiltext{3}{National Astronomical Observatories, Chinese
Academy of Sciences, Beijing 100012, China}

\begin{abstract}
We show that redshift-space distortions of galaxy correlations have a
strong effect on correlation functions with distinct, localized
features, like the signature of the baryon acoustic oscillations
(BAO).  Near the line of sight, the features become sharper as a
result of redshift-space distortions.  We demonstrate this effect by
measuring the correlation function in Gaussian simulations and the
Millennium Simulation.  We also analyze the SDSS DR7 main-galaxy
sample (MGS), splitting the sample into slices $2.5\degr$ on the sky
in various rotations.  Measuring 2D correlation functions in each
slice, we do see a sharp bump along the line of sight.  Using
Mexican-hat wavelets, we localize it to $(110\pm 10)\hMpc$.  Averaging
only along the line of sight, we estimate its significance at a
particular wavelet scale and location at 2.2$\sigma$.  In a flat
angular weighting in the $(\pi,r_p)$ coordinate system, the noise
level is suppressed, pushing the bump's significance to 4$\sigma$.  We
estimate that there is about a 0.2\% chance of getting such a signal
anywhere in the vicinity of the BAO scale from a power spectrum
lacking a BAO feature.  However, these estimates of the significances
make some use of idealized Gaussian simulations, and thus are likely a
bit optimistic.

\end{abstract}


\section{Introduction}
In a wide range of cosmological models, acoustic oscillations were
generated due to competition between the pressure exerted by the
photons and the gravitational collapse of perturbations in the density
of baryons in the relativistic baryon-photon plasma of the early
universe prior to the epoch of recombination
\citep{s68,py70,sz70,be84,be87,h89}. These are imprinted not only on
the temperature power spectrum of the cosmic microwave background
(CMB), as detected convincingly for the first time around the turn of
the millennium \citep{deb00, h00}, but also on the power spectrum of
matter and galaxies.  The Baryon Acoustic Oscillation (BAO) feature
was first convincingly detected in the correlation function of
Luminous Red Galaxies (LRGs) in the Sloan Digital Sky Survey (SDSS) by
\citet{e05}. Further evidence for the BAO has been found subsequently,
in the SDSS and other galaxy surveys \citep{c05, pad07, p07,p10}.

Some of these papers used correlation functions, others used power
spectra. Correlation functions have both advantages and disadvantages
over power spectra. For example, when there are quasi-harmonic
features, like the BAO, forming a sequence of peaks in $k$-space, they
can add to a much stronger feature in the correlation function since
the harmonics all add to the same peak in $\xi$.  A disadvantage of
the correlation function is that unlike the power spectrum, it suffers
from substantial off-diagonal covariances, even on large scales
\citep[e.g.\ ][]{ham09}.  However, it has recently been shown that
even the power spectrum is not entirely immune from covariance effects
in the mildly nonlinear regime \citep{rh05,nsr06}.

Correlation functions of lower-dimensional subsets for a homogeneous
isotropic random field are identical to the one estimated from the
full three-dimensional one. This can also be extended to density
fluctuations with line-of-sight anisotropies caused by redshift space
distortions. Over the last two decades there were many redshift
surveys, which had pencil-beam- or slice-like geometries
\citep[e.g.][]{cfa,l96,beks,virmos,deep}.  Even wide angle surveys
like the 2dF \citep{c01} had many pencil-beams as outrigger fields for
a random sampling of a larger volume, while making optimal use of the
multi-fiber capabilities of the telescope over a limited field of
view. In any case, all these geometries should lead to the same
redshift space correlation function. However, there are subtle effects
due to the geometry entering the covariance of the correlation
function.

The statistical analysis of significance for correlation functions is
a notoriously difficult task \citep{kp91}, since the different bins of
the correlation function are always correlated to one another.  The
different modes in the power spectrum are more independent, as
long as the survey window is large enough, making estimation in many
ways easier \citep{t97}. For anisotropic windows the shape of the
independent ``grains'' in $k$-space can be very elongated, leading to
complications, like mixing modes over a wide range of scales. This is
particularly true for pencil-beams or slices, where power spectra have
rarely been used successfully to characterize structure.

Because the behavior of the BAO at low redshift only slightly departs
from the predictions of linear cosmological perturbation theory, the
BAO signature has recently attracted attention as a powerful
``standard cosmological ruler.''  It is a useful probe to explore the
nature of dark energy or large-distance modifications of gravity,
i.e.\ explanations of the observed accelerated expansion of the
Universe.

There is a recent controversy about the reality of a narrow peak at
the BAO scales along the line of sight (LOS) in the redshift-space
power spectrum of the SDSS LRG sample \citep{g09}.  Gravitational
lensing is mentioned as a possible origin of this signal, but this
explanation has been disputed \citep{me09}.  \citet{kazin10}, using a
suite of realistic LRG mock catalogs, conclude that the LOS peak is
likely a statistical fluke.  Recently, \citep{cabre10} reached a
similar conclusion about the significance of BAO detections, but
argued that even a modestly significant peak can be used to constrain
cosmological parameters, under the reasonable assumption that the
galaxy power spectrum in the Universe contains a BAO feature.

In these papers, the statistical significance of a BAO feature is
assessed by comparing data to models without BAO; for example, if a
correlation function is better-fit by such a ``no-wiggle'' model, the
features are judged to be spurious.  Here, we test a different
technique, detecting peaks with a Mexican-hat wavelet.  Although it
seems more forgiving of modest BAO-like peaks than the standard
approach, we show that it still has much BAO-location constraining
power

In this paper, we revisit the linear theory of redshift space
distortions, and show that for power spectra with distinct spectral
features, redshift space distortions will have a characteristic
sharpening effect on these, especially along the line of sight,
even in linear theory. 
In Section 2, we demonstrate how different 
geometric sampling strategies of a given survey volume can
lead to strong, but quite different covariances in the resulting
correlation functions.  
In Section 3, we describe our wavelet-based technique for BAO peak
detection, and measure its statistical and systematic uncertainties
using simple Gaussian simulations.
In Section 4, we analyze a sample of galaxies from the Millennium
simulation \citep[MS][]{mill} using the wavelet technique.
In Section 5, we build a sample of SDSS DR7 \citep{dr7} main-galaxy
sample galaxies, and analyze the LOS and 2D correlation functions
using a methodology similar to the simulations: we subdivide the
sample into many thin slices, compute the 2D redshift space
correlation function and calculate the average.
In Section 6, we apply the wavelet peak-finding formalism to the SDSS results.
Finally, in the conclusion, we discuss our findings.

\section{Redshift Space Correlations in Linear Theory}
\label{sec:theory}
\subsection{Amplifying the BAO Features}

The seminal paper of \citet{k87} laid out the framework to compute the
redshift space distortions of power spectra, in the plane-parallel
limit, when the two galaxies are very close to the same LOS. Later,
\citet{h92}, and \citet{hc96} worked out explicit expressions for the
correlation function. \citet{ht95} considered the problem for all-sky
redshift surveys using a spherical-harmonic analysis.

\citet[][hereafter SML98]{sml98} used bipolar spherical harmonics to 
extend the previous calculations of the correlation function to arbitrary 
angles between the two lines of sight, and have computed explicit expressions 
for the distorted correlation function in different directions. Their work 
was further extended by \citet{s04} and \citet{papais08}, using a slightly 
different coordinate system, and including contributions from a previously 
neglected term in the Jacobian of the real- to redshift-space mapping. 
\citet{scocc04} included various non-linear effects, in particular 
contributions from the nonlinear pairwise velocity distributions.

The basic papers laying out the theory of redshift-space distortions
were written in the 80's and 90's, when the generally accepted
assumption was that the cosmological dark matter power spectrum is
smooth. Here we would like to show that one of the effects of the
redshift space distortions on the correlation function is a sharpening
of ``bumpy'' features, in directions close to the LOS.

Following SML98, we can write the redshift space correlation function
as a function of the pairwise distance $r$, and the angle $\theta$
($\gamma$ in SML98), using the expression for the plane-parallel limit
(especially applicable near the LOS), as
\begin{eqnarray}
  \xi(r,\theta) &=& \left( 1+2\beta/3+\beta^2/5\right)\,\xi_0(r)\nonumber\\
 & &-\left(4\beta/3+4\beta^2/7\right)\,\xi_2(r) P_2(\cos \theta) \nonumber\\
 & &+\left(8\beta^2/35\right)\,\xi_4(r) P_4(\cos \theta),
 \label{eqn:pp}
\end{eqnarray} 
where $P_n(x)$ is the Legendre polynomial of order $n$, $\beta$ is the
usual redshift space distortion parameter, $\beta=\Omega_m^{0.6}/b$,
where $b$ is the bias factor, and $\xi_L(r)$ is the $L$-th spherical
Bessel transform of the isotropic, real-space power spectrum $P(k)$, 
\begin{eqnarray}
	\xi_L(r) = \frac{1}{2\pi^2}\int dk\/ k^2\/ j_L(kr) P(k).
\end{eqnarray} 
The three-dimensional angular average of Eq.\ (\ref{eqn:pp}) is
$\bar\xi_{\rm 3D}=(1+2\beta/3+\beta^2/5)\xi_0(r)$, using a
$\sin\theta\ d\,\theta$ weighting; only the isotropic $\xi_0$
contributes due to the orthogonality of the Legendre polynomials.
However, such a weighting suppresses contributions from close to the
LOS, where $\sin\theta=0$.  If $\xi(\pi,r_p)$ is averaged with a flat
(without the $\sin\theta$ factor) angular weighting, appropriate for a
2D sample, there are also contributions from $\xi_2$ and $\xi_4$:
\begin{eqnarray}
  \bar\xi_{\rm 2D}(r) &=& \left(1+2\beta/3+\beta^2/5\right)\,\xi_0(r)
  -\left(\beta/3+\beta^2/7\right)\,\xi_2(r)\nonumber\\ 
&+& \left(9\beta^2/280\right)\,\xi_4(r).
 \label{eqn:flatxi024}
\end{eqnarray} 
 
\begin{figure}
  \begin{center}
    \includegraphics[scale=0.48]{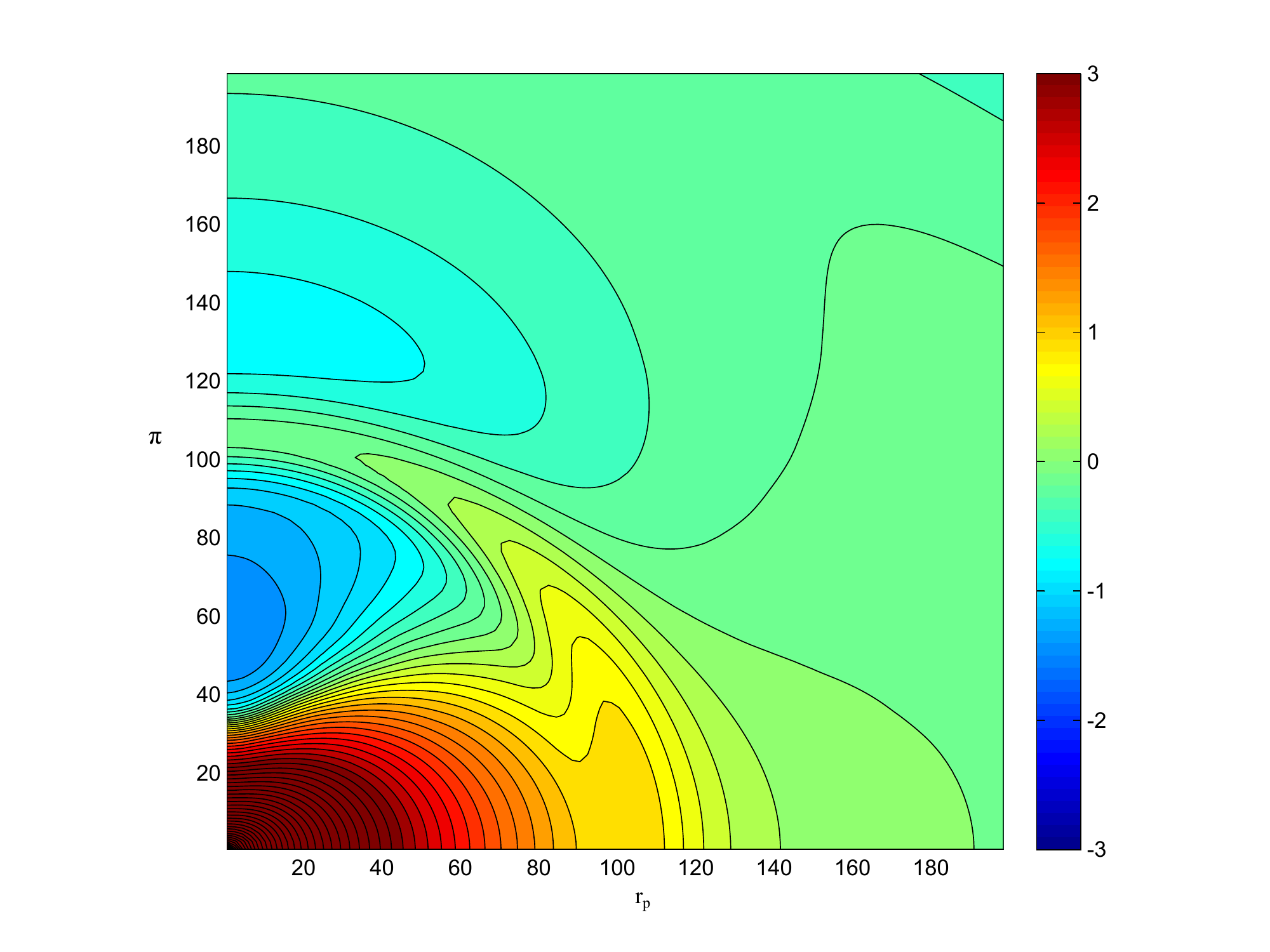}
  \end{center}
  \caption{The two-dimensional correlation function for a power
    spectrum with a BAO feature. Note how the bump gets sharper, but
    also lower, as it approaches the LOS. The function plotted is
    $\sinh^{-1}(300\xi(\pi,r_p))$, linear near zero, but logarithmic
    for high values of $\xi$.  Units on both axes are $\hMpc$.
	\vspace{0.1in}
 \label{fig:linear2D}
  }
\end{figure}

Figure \ref{fig:linear2D} shows contours of the linear redshift-space
correlation function $\xi(\pi,r_p)$ derived from a power spectrum with
a BAO bump at 107$\hMpc$.  In linear theory, the BAO feature sharpens
near the LOS.  Here $\pi$ and $r_p$ are galaxy separations projected
parallel and perpendicular to the LOS.  We use a $\sinh^{-1}(300\xi)$
transform for easy visualization for both large and small $\xi$.  The
$\sinh^{-1}(x)$ function equals $x$ for $x\ll 1$, and $\ln x$ for
$x\gg 1$.

\begin{figure}
  \begin{center}
    \includegraphics[scale=0.4]{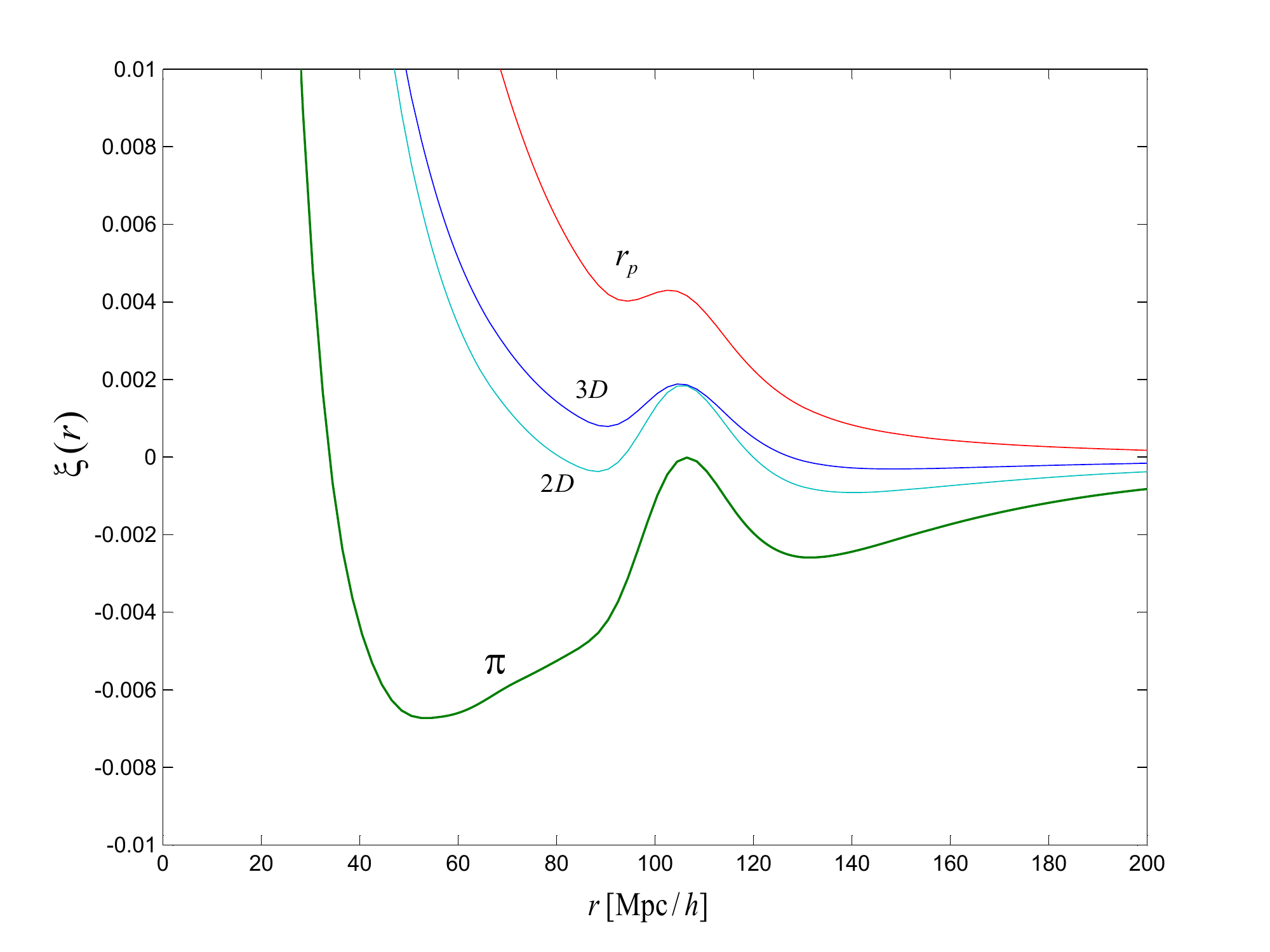}
    \includegraphics[scale=0.4]{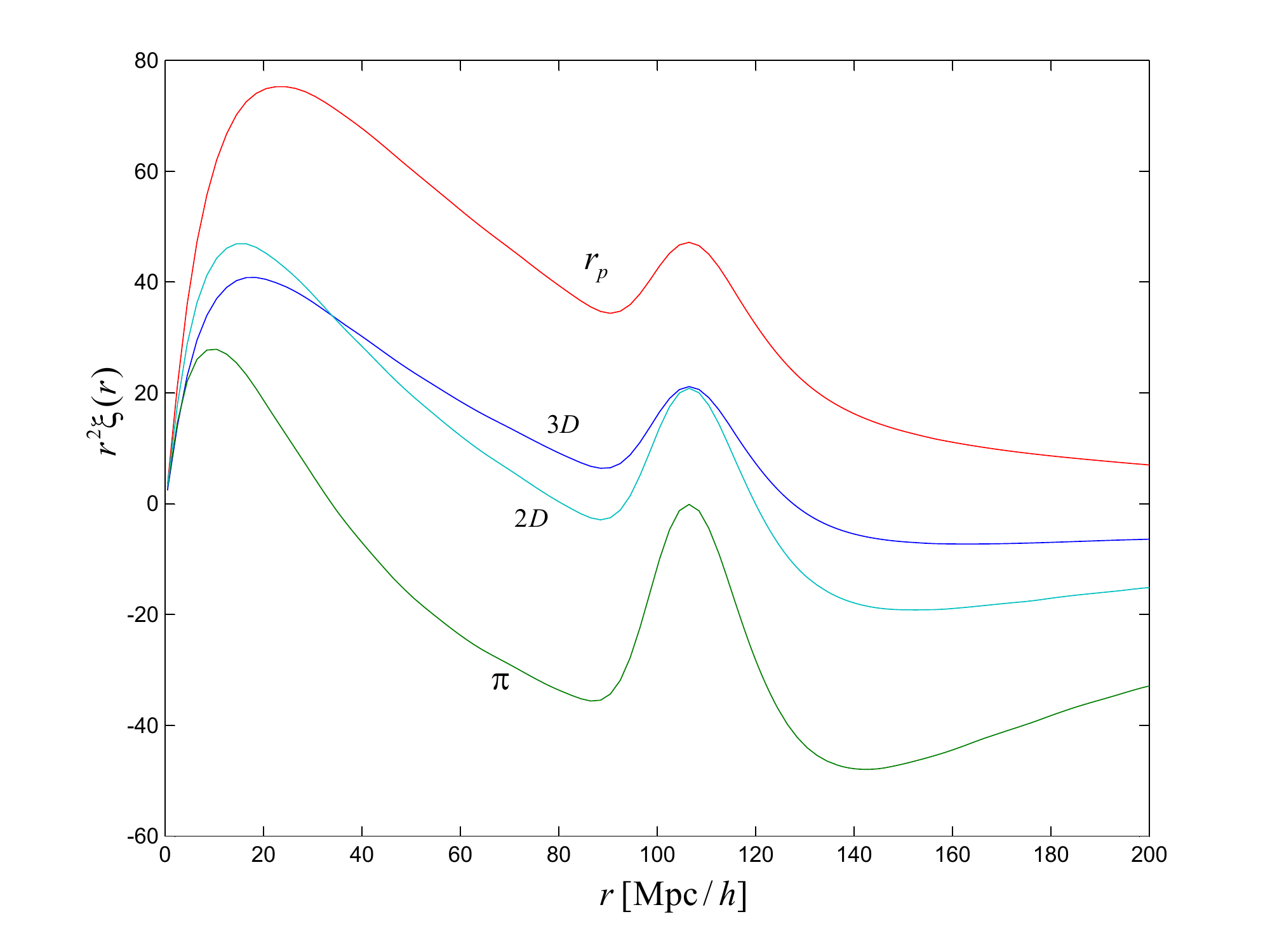}
  \end{center}
  \caption{The two-dimensional linear theory correlation function is
    shown along the LOS ($\pi$), along the transverse direction
    ($r_p$), together with two angular averages of the correlation
    function. The 3D curve uses a $\sin\theta$ weighting, where
    $\theta$ is the angle away from the LOS in the ($\pi,r_p$)
    coordinate system.  The 2D curve, which has a sharper bump, uses a
    flat (without the $\sin\theta$) weighting.  The lower panel shows
    the various $\xi$'s, after multiplying by $r^2$.
    \label{fig:xicutslinear}
  }
\end{figure}

The LOS correlation function, $\xi(\pi)$, can be written as
\begin{eqnarray}
\xi(\pi) &=& \left( 1+2\beta/3+\beta^2/5\right)\,\xi_0(\pi) - \nonumber\\
 & &\left(4\beta/3+4\beta^2/7\right)\,\xi_2(\pi) + \left(8\beta^2/35\right)\,\xi_4(\pi).
\end{eqnarray} 
The term independent of $\beta$ is just the isotropic, real-space
correlation function.  The term linear in $\beta$ can be written as
$(2\beta/3)[\xi_0(\pi)-2\xi_2(\pi)]$, which inside the $k$-integrals
behaves as $j_0(k_\pi)-2j_2(k_\pi)= -3j_0''(k_\pi)$, the second
derivative of the spherical Bessel function. Such operators were
explicitly shown in Eq.\ (4) of \citet{h92}. Applying a second
derivative to a Gaussian, and adding it to the function with a
positive weight will sharpen the peak, as seen in
Fig.\ \ref{fig:xicutslinear}. The sharpening is much weaker far from
the LOS, in the transverse ($r_p$) direction. The other effect of
redshift-space compression is an overall smooth shift towards negative
values along the LOS, and towards positive values in the transverse.

In Fig.\ \ref{fig:xicutslinear}, we also show the results if $\xi$ is
multiplied by $r^2$, in which case the peaks at different angles line
up in scale more obviously.  This could be because the slopes on which
the peaks find themselves at different angles are much more similar in
$r^2\xi$ than in $\xi$ itself.  This is an example illustrating how
using the absolute position of $\xi$'s local maximum, without regard
to the slope it is on, can produce a small bias.  A peak locator that
estimates the second derivative, as our wavelet transform does, is
insensitive to such slopes.

\begin{figure}
  \begin{center}
    \includegraphics[scale=0.43]{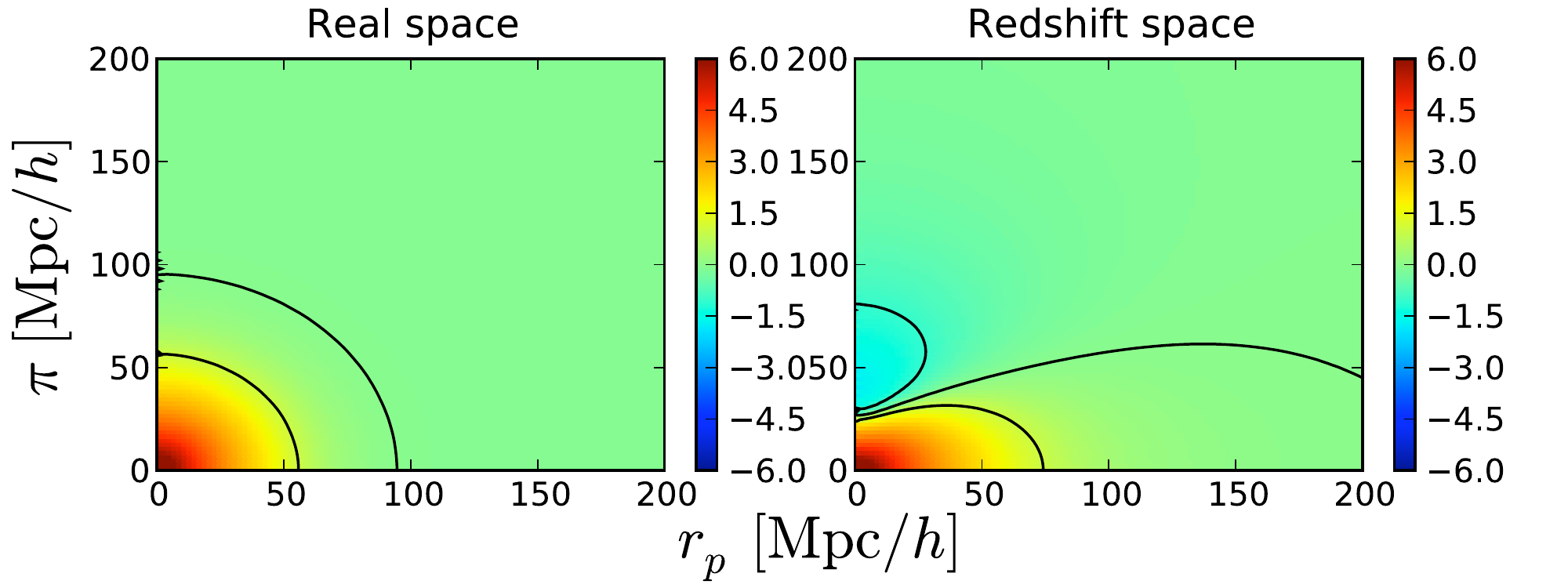}
  \end{center}
  \caption{$\sinh^{-1}(300\xi)$ from an \citet{ehu} linear no-wiggle
    power spectrum, with and without linear redshift-space
    distortions.  The contours are at -1, 0, and 1.
    \label{fig:xinowig}
  }
\end{figure}

Fig.\ \ref{fig:xinowig} shows an \citet{ehu} no-wiggle linear
$\xi(\pi,r_p)$ in real and redshift space.  Note that the trough along
the LOS at $\pi\approx 55\hMpc$ appears even in the no-wiggle case,
and thus seems unrelated to the BAO feature in linear theory.
However, as shown below in Fig.\ \ref{fig:xi1d}, the trough is highly
amplified in samples with full nonlinearities (in the SDSS and MS
samples); we suspect that this could be a result of non-linear infall
toward the BAO ridge.

For every pair of galaxies, the observer and the two galaxies define a
plane, and the whole problem is invariant under any rotation of this
plane around the observer, located at the origin. Thus the anisotropic
redshift space correlation function is inherently planar, and in 3D it
has an axial symmetry for rotations around the LOS. This means that
one can also estimate the same 2D correlation function from an
arbitrarily thin slice of the data (though if one goes too thin, shot
noise will swamp the signal). From the projection-slicing theorem,
described below, this is equivalent to first projecting the
three-dimensional redshift-space power spectrum down to two
dimensions, and then perform a two-dimensional inverse Fourier
transform. This has been noted in SML98. Computing the anisotropic
redshift space correlation function this way does not violate any of
the underlying symmetries of the problem. However, it impacts the
covariances of the estimated correlation function, as we will show.

\subsection{Projection-Slicing Theorem}

There is well-known theorem in signal processing and medical imaging, 
stating a relation between the Fourier transform of a lower dimensional
subset of a multidimensional scalar field and the original Fourier 
transform.

The projection-slicing theorem states that the Fourier transform of
the projection of an $N$-dimensional scalar function $f$ onto an
$m$-dimensional subspace is equal to an $m$-dimensional slice, going
through the origin of the $N$-dimensional Fourier transform of $f$.
For example, if a 3D $\delta$ is projected to 2D along the $z$
direction, the Fourier transform of the 2D projection will be on the
$k_z$ plane of the 3D Fourier transform of $\delta$.  Symbolically,
\begin{equation}
	F_m P_m = S_m F_N,
\end{equation}
where $F_k$ denotes the $k$-dimensional Fourier transform, $P_m$
denotes a projection onto the $m$-dimensional subspace, and $S_m$
denotes slicing of the $m$-dimensional subspace.  This technique will
be useful for the calculations below.

We can compute the redshift space correlation function in several
different ways. We can take the full 3D redshift space correlation
function, and perform the axial averaging using the LOS as the
symmetry axis to get the 2D $\xi(\pi, r_p)$. Alternatively, we can
compute the 2D correlation function directly from each slice in an
ensemble, and then perform the averaging in 2D.

In the case of LOS correlations, one can do the either of the above 
procedures, then take the LOS, or one can even directly extract all possible 
pencil-beams from the volume and just compute the 1D correlation function. 
In the extreme, near the LOS, all three cases will contain the same galaxy pairs,
nevertheless the covariances between the correlation function bins 
estimated in different ways will be somewhat different due to lateral correlations
between neighboring slices and pencil-beams. In the following subsections we
will illustrate how to compute the covariance of the correlation functions, at 
least in the linear limit, where the three results are quite intuitive, and 
show the subtle differences between the outcomes. Furthermore, even though the 
calculation below is only for the LOS correlation function, it can be easily 
extended for the covariance of the full 2D redshift space correlation function.

\subsection{Estimating the covariance of $\xi_{LOS}(r)$ from pencil-beams}

We ignore the effects of shot noise, and only estimate the variance
and covariance of the correlation function from the power
spectrum. Consider first a single pencil-beam, aligned with the
$z$-axis, drawn randomly from a cubic volume with periodic boundary
conditions, like the case of the N-body simulations we analyzed in
this paper. The overdensity is measured in $N$ discrete cells, along
each axis of the cube. The estimator for the LOS correlation function
is
\begin{equation}
	\xh_1(r) = \frac{1}{N}\sum_i \delta(r_i) \delta(r_i+r),
\end{equation}
where $\delta(r_i)$ is the dimensionless overdensity in the $i$th cell
along the LOS.  The expectation value $\avg{\xh_1(r)}= \xi_1(r)$.

We can compute the expectation value
\begin{equation}
	\avg{\xh_1(r) \xh_1(r')} = \frac{1}{N^2} \sum_{i,j} 
	\avg{\delta(r_i) \delta(r_i+r) \delta(r_{j}) \delta(r_{j}+r')}.
\end{equation}
and the covariance
\begin{equation}
	C_1(r,r') = \avg{\xi_1(r) \xi_1(r')} - \avg{\xi_1(r)}\avg{\xi_1(r')}.
\end{equation}
If we stay in the Gaussian (linear) limit, there are no higher order, 
irreducible contributions, thus the expectation value of the product
of four overdensities can be factored as 
\begin{equation}
	\avg{\delta_1 \delta_2 \delta_3 \delta_4} =
	\avg{\delta_1 \delta_2}\avg{\delta_3 \delta_4} +
	\avg{\delta_1 \delta_3}\avg{\delta_2 \delta_4} +
	\avg{\delta_1 \delta_4}\avg{\delta_2 \delta_3}.
\end{equation}
With $r'=r_{i}+s$, we can write the expectation value as
\begin{eqnarray}
	\avg{ \xh(r)\xh(r')} &=& \frac{1}{N^2} \sum_{i,s} \nonumber \\
	&\big[&\avg{\delta(r_i)\delta(r_i+r)}\avg{\delta(r_i+s)\delta(r_i+s+r')} \nonumber \\
	&+& \avg{\delta(r_i)\delta(r_i+s)}    \avg{\delta(r_i+r)\delta(r_i+s+r')} \nonumber \\
	&+& \avg{\delta(r_i)\delta(r_i+s+r')} \avg{\delta(r_i+r)\delta(r_i+s)} \big] 
\end{eqnarray}
After the summation over the index $i$, we obtain
\begin{eqnarray}
	\avg{\xh(r) \xh(r')} &&=  \xi_1(r)\xi_1(r') + \nonumber \\
	\frac{1}{N} \sum_{s} &&
	\Big[\xi_1(s)\xi_1(s+r'-r)+\xi_1(s+r')\xi_1(s-r) \Big].
\end{eqnarray}
The first term is just the trivial product of the expectation values.
The next terms are the correlations of the correlation function,
which we will denote as $Z_1$, a symmetric function of its argument, as
\begin{equation}
	Z_1(z) = \frac{1}{N} \sum_{s} \xi_1(s)\xi_1(s+z).
\end{equation}
We can compute $Z_1$ in the limit of infinitesimal cells, using the Fourier
transform of the one dimensional $\xi_1(z)$, denoted as $\pi_1(k_z)$.
Due to the projection-slicing theorem, the one-dimensional power spectrum
corresponding to the correlation function along the pencil-beam is the 
projection of the three-dimensional power spectrum $P$, anisotropic 
due to the redshift space distortions along the $z$-axis:
\begin{equation}
	\pi_1(k_z) = \frac{1}{(2\pi)^2} \int dk_x dk_y P(k_x,k_y,k_z),
\end{equation}
and 
\begin{equation}
	Z_1(z)=  \frac{1}{2\pi} \int dk_z e^{i k_z z} |\pi_1(k_z)|^2.
\end{equation}
In summary, we can write the covariance of the 1D correlation function as 
estimated from a single pencil-beam as
\begin{equation}
	C_1(r,r') = Z_1(r-r') + Z_1(r+r').
\end{equation}

\subsection{The covariance of $\xi_{LOS}(z)$ for independent slices}

Next, we randomly select a thin density slice from a cubic volume, and 
estimate the two dimensional correlation function. We essentially 
repeat the derivation above, except we have two indices: $i$ is along the 
LOS, and $j$ is along the slice:
\begin{equation}
	\xh_2(s,r) = \frac{1}{N^2}\sum_{j,i} \delta(r_j,r_i) \delta(r_j+s,r_i+r).
\end{equation}
The LOS correlation function is the special case with $s=0$. Following 
the previous calculation, we obtain the equivalent 2D covariance as
\begin{equation}
	C_2(r,r') = Z_2(0,r-r') + Z_2(0,r+r').
\end{equation}
with $Z_2$ now related to $\pi_2(k_y,k_z)$, the 2D projection of the 3D 
power spectrum:
\begin{equation}
	\pi_2(k_y, k_z) = \frac{1}{2\pi} \int dk_z P(k_x, k_y, k_z),
\end{equation}
\begin{equation}
	Z_2(s,r) = \frac{1}{(2\pi)^2} \int dk_y dk_z e^{i (k_y s + k_z r)} 
		|\pi_2(k_y, k_z)|^2.
\end{equation}
Along the LOS, $Z_2(0,r)$ can be written as:
\begin{equation}
	Z_2(0,r) = \frac{1}{2\pi} \int dk_z e^{i k_z r} 
	  \int \frac{dk_y}{2\pi} |\pi_2(k_y, k_z)|^2
\end{equation}
This is the projection-slicing theorem at work again: we are 'slicing' the
2D super-correlation function at $s=0$, therefore, we need to project its
power spectrum down to the $z$-axis. This is however different from the previous
result, there the projection has happened for the power spectrum, before the
square was taken. Here we performed one projection before and one after taking 
the square of the power spectrum.

\subsection{The covariance from the average of slices}

We repeat the calculation of the LOS correlation function as above,
but now we average over the whole three-dimensional cube. We can
perform this by subdividing the volume into a set of adjacent slices,
computing the correlation function of each slice, then averaging them
over the whole set.  The derivation is quite similar, except now we
have a 3D estimator of the LOS correlation function. We will just
present the final result:
\begin{equation}
	\xh_3(0,0,r) = \frac{1}{N^3}\sum_{k,j,i} \delta(r_k,r_j,r_i) \delta(r_k,r_j,r_i+r)
\end{equation}
\begin{equation}
	C_3(r,r') = Z_3(0,0,r-r') + Z_3(0,0,r+r'),
\end{equation}
with
\begin{equation}
	Z_3(0,0,r) = \frac{1}{2\pi} \int dk_z e^{i k_z r} 
	  \int \frac{dk_x dk_y}{(2\pi)^2} |P(k_x, k_y, k_z)|^2 
\end{equation}

Here all projection takes place after squaring the anisotropic 3D
power spectrum.  Even without evaluating any of the integrals one can
see that the covariances will be the largest in this, third case,
since the quadratic mean is always larger than the arithmetic one. So
if at a given $k_z$ we project the square of the 3D power spectrum, we
always get a larger number if some of the projection takes place
before taking the square.

\subsection{Effect of strong covariances on the estimated errors}

We can write a linear transform of correlation function values in bins
as $a = \sum_i u_i x_i$, where $u_i$ are the coefficients of the
linear transformation, and $x_i$ are the binned values of the 1D
correlation function. Assuming that the transform has a finite
support, the sum is over a small, finite number of bins. The
expectation value of $a$ can be written as
\begin{equation}
       \langle a \rangle = \sum_i u_i \langle x_i \rangle
\end{equation}
For a wavelet transform, the vector $u$ contains some positive and negative 
coefficients, so that their sum is equal to 0. The expectation value of $a^2$ is
\begin{equation}
       \langle a^2 \rangle = \sum_{i,j}  u_i u_j \langle x_i x_j\rangle,
\end{equation}
and the variance is
\begin{equation}
{\rm Var}(a) =\langle a^2 \rangle - \langle a \rangle^2  
= \sum_{i,j}  u_i u_j [ \langle x_i x_j\rangle -  
\langle x_i \rangle \langle x_j\rangle ].
\end{equation}

Consider a simple wavelet of length 3: $u=(-0.5, 1, -0.5)$.  The
covariance of the correlation function is dominated by the $Z(r-r')$
term. Since this is a symmetric function of its argument, it is
reasonable to approximate it with an inverse parabola $Z(r) = (1-a
r^2)$, taking also unit diagonal variance. Expanding the covariance
matrix up to second order in $r$
\begin{equation}
	\left( {
  \begin{array}{ccc}
	1    & 1-a & 1-4a\\
	1-a  & 1   & 1-a \\
	1-4a & 1-a & 1   \\
   \end{array}
	} \right)
\end{equation}
then the variance is given by $u^T Z u = 0$.  The covariances of the
LOS correlation function are quite high.  Thus, applying a similar,
but much wider wavelet can potentially reduce the variance
considerably compared to the nominal diagonal variances, although it
will never go to zero as in this extreme example. This Section only
serves as an illustration of the effect of the strong covariances, and
the importance of the particulars of the sampling strategy chosen.

Let us describe what this means in simple terms. Consider first a set
of independent, uncorrelated slices randomly drawn from a infinite
sample, and denote the variance of a wavelet coefficient at some scale
over this ensemble as $\sigma_2^2$. This variance will be already
smaller than the variance of the individual bin values of the
correlation function, since the covariance among the bins
[$C_2(z,z')$] is reducing the wavelet variance.

Next, let us draw a set of adjacent slices from a coherent cubical
volume, and compute the average wavelet coefficient over this
sample. The covariance of the estimator $\xi_3$ now becomes much
stronger, $C_3(z,z')$; thus, the variance of the wavelet coefficient
will be much smaller.

If we took the average over the same number of slices from the
independent ensemble, the variance would be reduced by $\sqrt{N_{\rm
    slices}}$, due to the central limit theorem. For $\xi_3$, though,
estimating the variance among the slices assuming they are independent
will be an underestimate of the true variance.  But $\sqrt{N}$ can be
used as a definite bound on the expected variance reduction from the
slice-averaging process. At the end of Section
\ref{sec:correlated_slices} we demonstrate this behavior using Gaussian
simulations.

\section{Quantifying the Redshift Space Features}

\subsection{Peak Location from a Mexican Hat Wavelet}

We measure the sharpness of peaks in the correlation function using a
Mexican hat wavelet, similar in spirit to the (differently shaped)
wavelet technique proposed by \citet{xx10}.  A transform using a
Mexican hat wavelet, a second derivative of a Gaussian, provides a
measurement of (minus) the second derivative of a function, estimated
over the scale radius of the wavelet.  Before discussing the Mexican
hat itself, we describe estimators of the zeroth and first derivatives
of a function, also using derivatives of a Gaussian.  These estimators
will turn out to be useful in visualizing the sharpness of the BAO
bump as a function of angle.

Define a Gaussian of scale radius $s$, and normalized to equal 1 at
its center (bump position $r_b$), as
\begin{equation}
  G_0(r_b,s;r)=\exp\left[-(r-r_b)^2/2s^2\right].
  \label{eqn:g0}
\end{equation}
(The arguments before the semicolon are really parameters of the
function itself, which acts on parameters after the semicolon.)  An
estimate of the mean of $\xi(\pi,r_p)=\xi(r,\theta)$ (i.e.\ the zeroth
derivative) within a radius $s$ of $r_b$ is given by the average
$\avg{G_0(r_b,s)\,\xi}\equiv\int G_0(r_b,s;r)\xi(r,\theta)\,dr/\int
G_0(r_b,s;r)\,dr$.  Here $\theta$ is the angle away from the LOS.

To estimate the radial first derivative within a radius $s$ of $r_b$,
we use the following, where $\rho\equiv(r-r_b)/s$:
\begin{equation}
  G_1(r_b,s;r)=\rho G_0(r_b,s;r),
  \label{eqn:g1}
\end{equation}
\begin{equation}
  \avg{G_1(r_b,s)\xi}=\frac{\int \rho G_0(r_b,s;r)\xi(r,\theta)\,dr}
{\int \rho^2 G_0(r_b,s;r) dr}.
\end{equation}
We use $\avg{}$ to denote a generalized average: in the denominator,
the terms multiplying $G_0$ in the integrand are squared.

An estimate of the second derivative is $\avg{G_2(r,b)\xi}$, with
the Mexican hat wavelet,
\begin{equation}
  G_2(r_b,s;r)=\left(1-\rho^2\right) G_0(r_b,s;r).
  \label{eqn:g2}
\end{equation}
The wavelet transform is given by
\begin{equation}
	\avg{G_2(r_b,s)\xi}=\frac{\int \rho G_2(r_b,s;r)\xi(r,\theta)\,dr}
{\int \rho^2 G_2(r_b,s;r) dr}.
\end{equation}
Numerically, what we do is to evaluate and integrate the numerator on
the same $\pi,r_p$ grid as $\xi$ is measured, multiplying the
integrand by $1/r$ to take out the $r$ weighting from the 2D
integration.  Note that whenever we use the transform, we divide means
by standard deviations, so in fact the normalization (in the
denominator) drops out.

We note that in practice, we analyze correlation functions that are
undefined for $r<0$, and while $\int_{-\infty}^\infty G_2(r_b,s;r) =
0$, $\int_0^\infty G_2(r_b,s;r) \ne 0$.  However, this problem is
negligible if $r_b \gtrsim 3s$, and we confine our analysis to such
cases.

\subsection{Flattening transformation with background subtraction}
\label{sec:flattening}

The redshift-space $\xi(r,\theta)$ has a strong quadrupole anisotropy,
making it difficult to see the strength of the peak at different
angles. To make a visual detection of the peak easier, we flatten
$\xi$ at a given radius $r_b$ by estimating, and then subtracting, the
zeroth and first derivatives with respect to both radius and angle.
This is the background subtracted correlation function, shown in Fig.
\ref{fig:flattening}. The kernels $G_{02}$ and $G_{12}$ are sensitive
to the angular quadrupole:
\begin{equation}
  G_{02}(r_b,s;r,\theta)=q_2(\theta)\,G_0(r_b,s;r);
  \label{eqn:g0mu}
\end{equation}
\begin{equation}
  G_{12}(r_b,s;r,\theta)=\rho q_2(\theta)\,G_0(r_b,s;r).
  \label{eqn:g1mu}
\end{equation}
Here $q_2(\theta)\equiv \cos^2\theta - 1/2$, designed to be zero at $45\degr$.

The following flattening transformation enhances the visual contrast
of features within a radial distance $s$ of $r_b$, essentially
removing the constant and linear terms of a Taylor expansion of $\xi$
in $r$ and $q_2$.
\begin{eqnarray}
  \bar{\xi}(r_b,s;\pi,r_p)&=&\xi(r,\theta)- \big[ \avg{G_0\xi}
	+q_2(\theta)\,\avgnol{G_{02}\xi}+ \nonumber \\
  & &\rho \avg{G_1\,\xi} + \rho q_2(\theta)\, \avgnol{G_{12}\xi}\big]. \nonumber\\
\end{eqnarray}
Importantly, though, the Mexican hat is insensitive to linear and
constant terms in $r$, so the flattening transformation leaves the
Mexican hat transform unchanged for any $(r_b,s)$.  Thus, the
flattening transformation helps one to see bump-like features
visually, but does not affect the quantitative assessment of them.
When we plot the flattened correlation functions below, we multiply
$\bar{\xi}(r_b,s;\pi,r_p)$ by $G_0(r_b,s)$ to emphasize the region to
which the wavelet is most sensitive.

\subsection{Wavelet transforms of Gaussian fields}
\label{sec:wavgauss}
Our method is to estimate the BAO scale by finding the peak in the
signal-to-noise ratio S/N of the wavelet transform.  At each
$(r_b,s)$, this S/N is the mean divided by the standard deviation,
over samples $i$, of the wavelet transform
$\avg{G_2(r_b,s)\xi_i(r,\theta)}$.

We first investigate the statistics and systematics of the BAO wavelet
estimator using Gaussian simulations, with mode amplitudes drawn from
a Rayleigh distribution about the mean linear power spectrum, and
using random phases.  The linear power spectrum is from
\camb\ \citep{camb}, using the same cosmological parameters as the
Millennium Simulation.  The Gaussian simulations have a box size of
768$\hMpc$, and a cell size of 2$\hMpc$.  Linear redshift distortions
are put on the density fields using Eq.\ (\ref{eqn:linearz}).  We
generated 1400 Gaussian simulations for each case below.

\begin{figure}
  \begin{center}
  \includegraphics[scale=0.45]{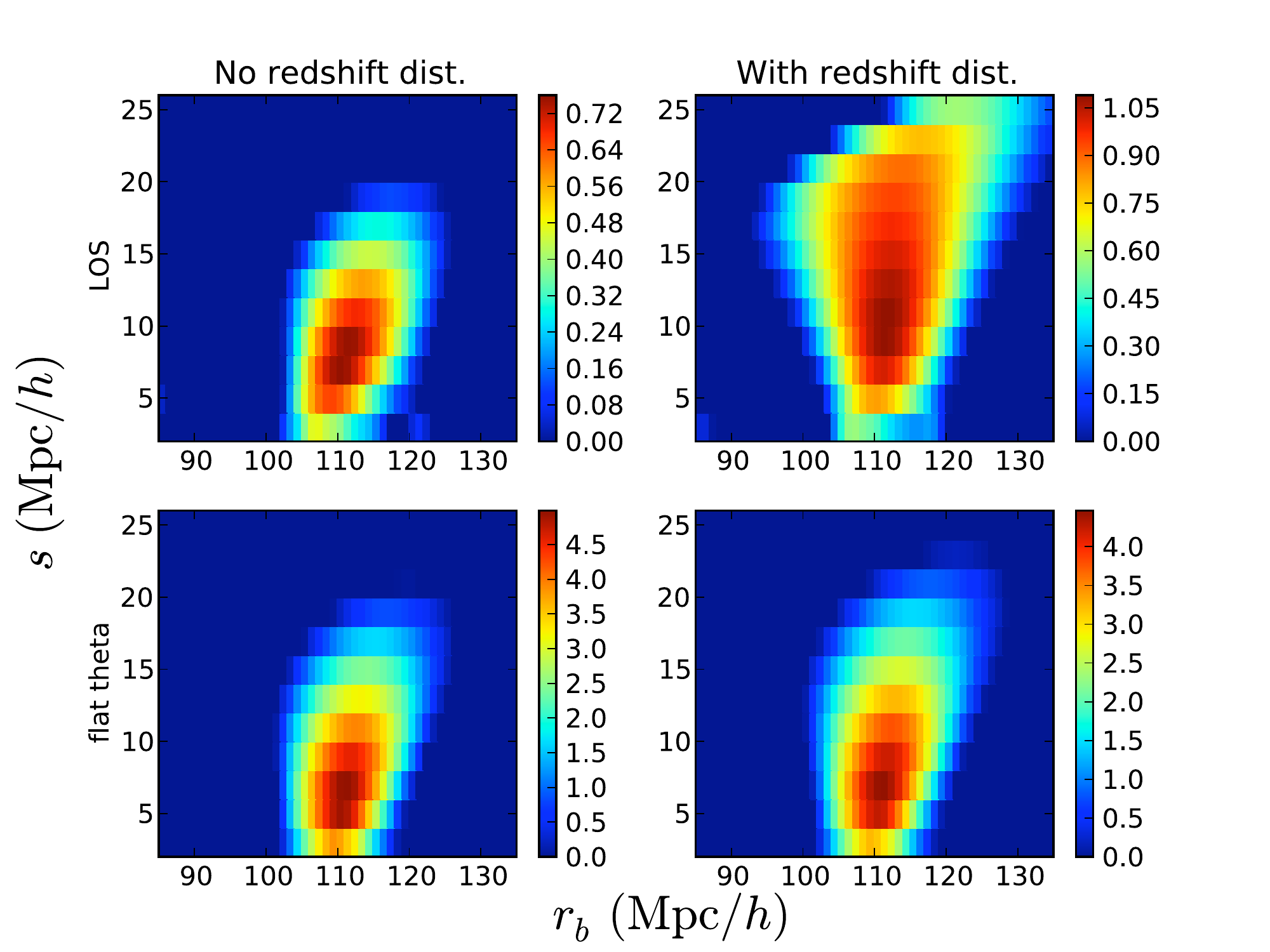}
  \end{center}
\caption{The signal-to-noise ratio of the wavelet transform
    $\avg{G_2(r_b,s)\xi(\pi,r_p)}$ applied to the full 3D correlation
    function of Gaussian realizations of volume (768$\hMpc$)$^3$,
    with and without linear redshift distortions applied.  The S/N
    assumes that a single box volume is being analyzed.
    \label{fig:gauss}
  }
\end{figure}

\begin{figure}
  \begin{center}
  \includegraphics[scale=0.45]{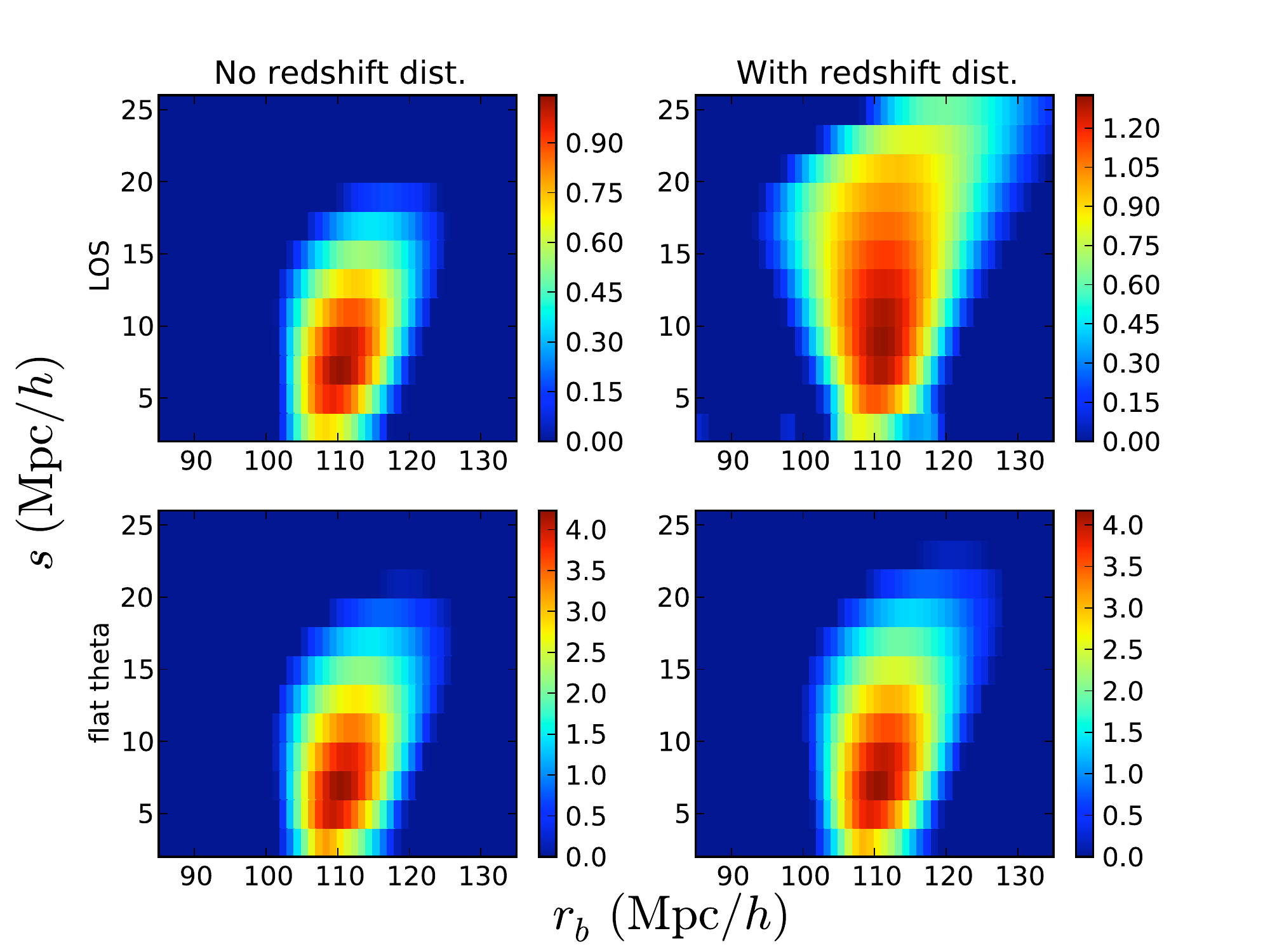}
  \end{center}
\caption{Same as Fig.\ \ref{fig:gauss}, for averaged 2D slice
    correlation functions.  The S/N assumes that a full box volume is
    being analyzed.
    \label{fig:gauss_slice16}
  }
\end{figure}

Figs.\ \ref{fig:gauss} and \ref{fig:gauss_slice16} show the wavelet
S/N from this ensemble of simulations, with and without slicing.
Linear redshift distortions are induced along the LOS $x$-axis, and
two slicings are taken, along the $y$ and $z$ axes.  The slice
thickness used is $16\hMpc$, which is the physical slice thickness of
the SDSS slices at a distance of about 370$\hMpc$.  

Each 4-plot figure shows the S/N in real and redshift space, and
weighting $\xi$ with two different angular weightings in the
$(\pi,r_p)$ plane: along the LOS (within $6\degr$ of the LOS,
$\mu^2=\cos^2 \theta > 0.99$); and with flat weighting in $\theta$.
In no case do we include the $(\sin \theta)$ weighting typically used
in angular averages of $\xi$ for 3D samples.  Such a weighting would
be inappropriate for the 2D slices that we consider.  Also, we do not
wish to kill the signal along the LOS.

In all cases, the S/N quoted is that expected from a single simulation
box, i.e.\ it is the mean divided by the standard deviation of the
wavelet transform, measured over the 1400 simulation boxes.  Note that
along the LOS, the S/N is enhanced for the averages of 2D slices.
This is even while missing some diagonal modes in the box, since we
use only 2 slicing orientations (along axes).  Most of the modes very
close to the LOS are picked up in at least one of the two
orientations, but in the flat weighting, many diagonal modes are
missing.  This is likely why, with flat weighting, the S/N degrades
with the slicings.

\begin{figure}
  \begin{center}
  \includegraphics[scale=0.45]{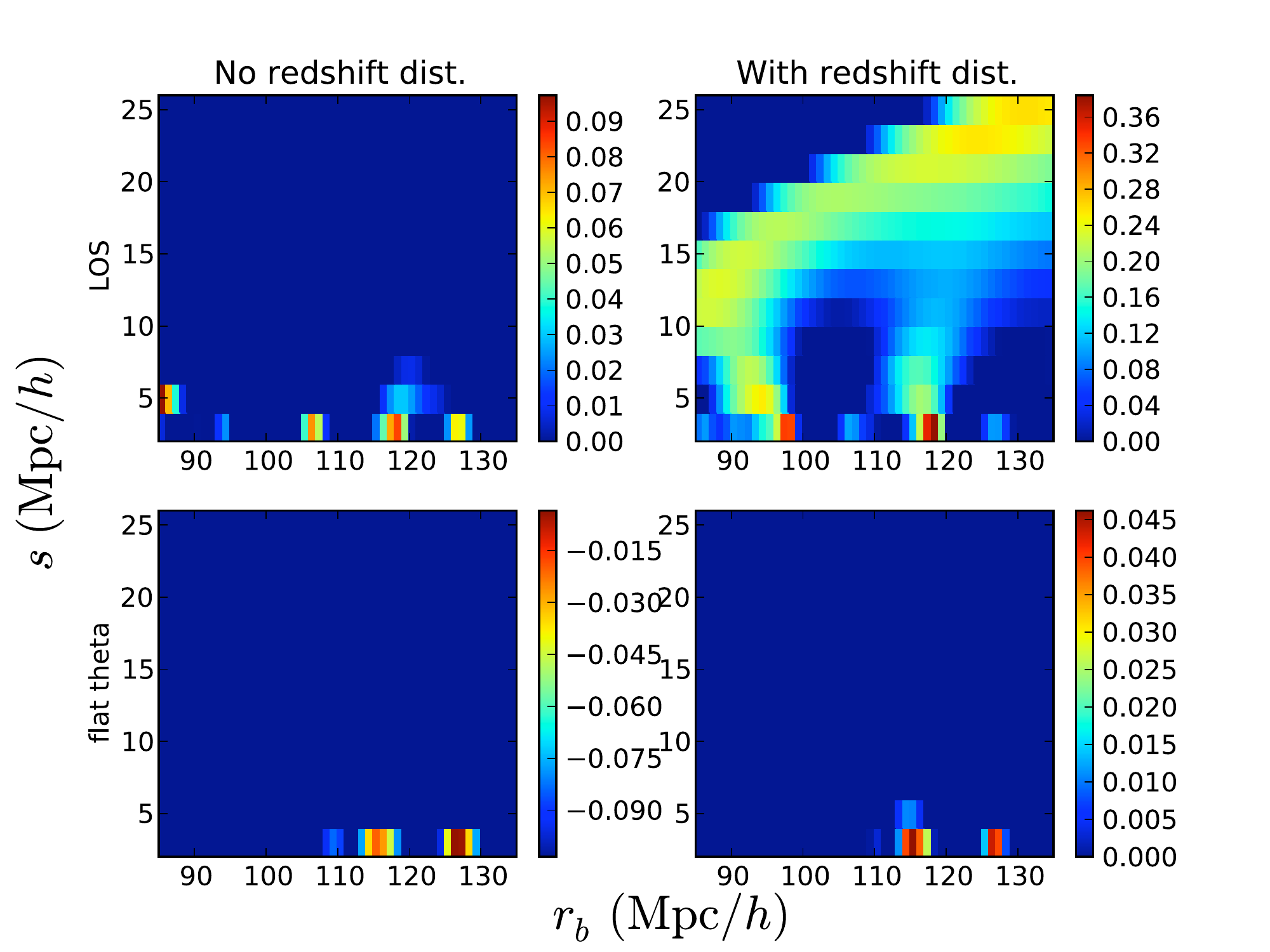}
  \end{center}
\caption{Same as Fig.\ \ref{fig:gauss}, except where the Gaussian
    simulations are generated using a no-wiggle power spectrum.
    \label{fig:gauss_ehu}
  }
\end{figure}

\begin{figure}
  \begin{center}
  \includegraphics[scale=0.45]{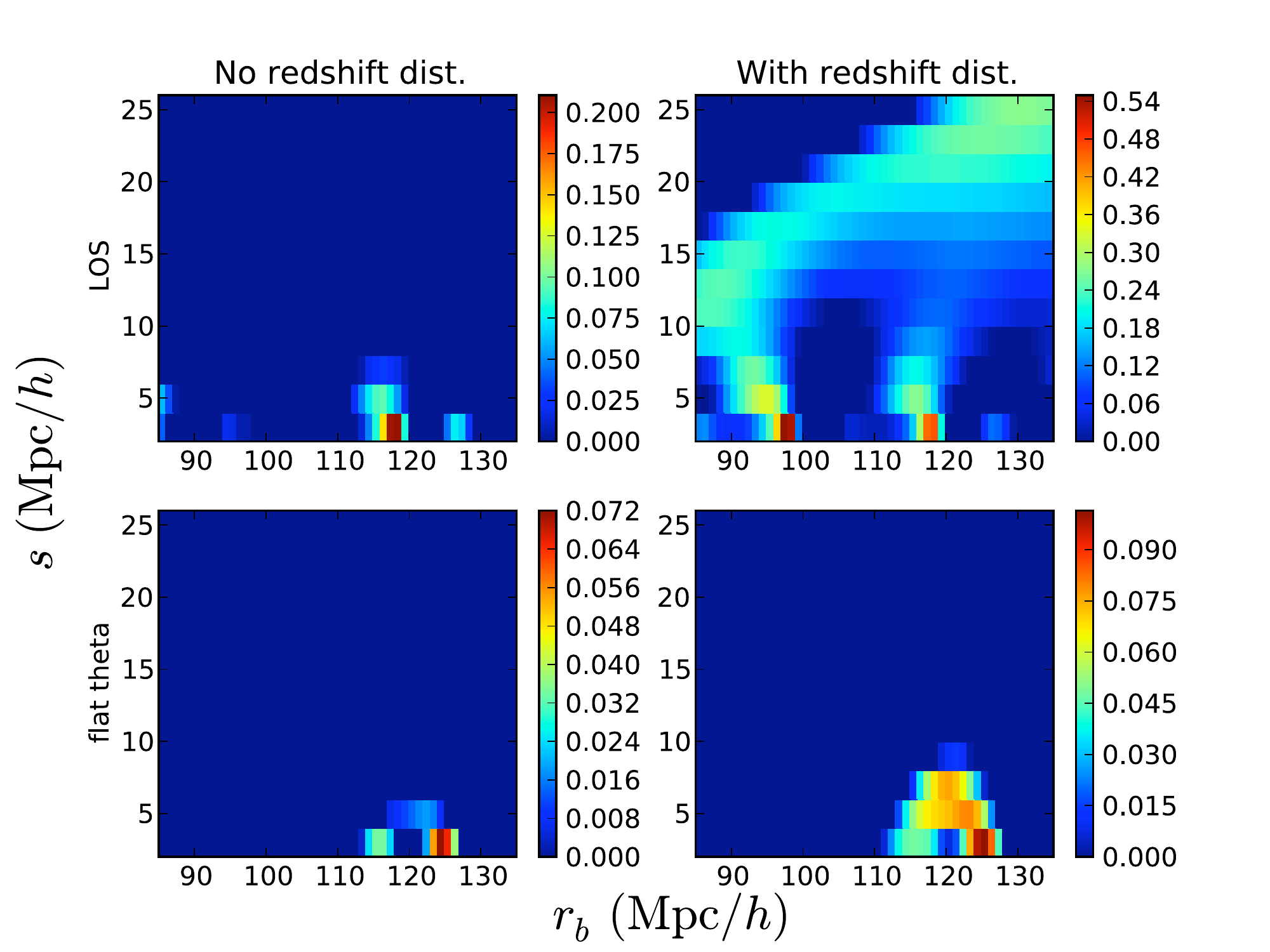}
  \end{center}
\caption{Same as Fig.\ \ref{fig:gauss}, for averaged 2D slice
    correlation functions, and where the Gaussian simulations are
    generated using a no-wiggle power spectrum.
    \label{fig:gauss_ehu_slice16}
  }
\end{figure}

Figs.\ \ref{fig:gauss_ehu} and \ref{fig:gauss_ehu_slice16} show the
analogs of the previous two figures, using no-wiggle power spectra.
The features detected in the no-wiggle case are generally small in
amplitude.  However, there is a ridge with S/N $\approx 0.24$ reaching
from $(r_b=75\hMpc,s=15\hMpc)$ to increasing $(r_b,s)$ along the LOS,
for the redshift-distorted $\xi_{\rm LOS}$.  This ridge is likely from
where the dip at $\pi\approx 55\hMpc$ flattens out.  This is a region
where the second derivative is substantially negative, but where the
first derivative is positive (thus it is not, strictly speaking, a
peak).  There are also peaks at $s=2\hMpc$ with fairly high
significance.  These are excluded from the peak-finding below, since
they lie along an edge of the searched prior in $s$.


\begin{figure}
  \begin{center}
  \includegraphics[scale=0.43]{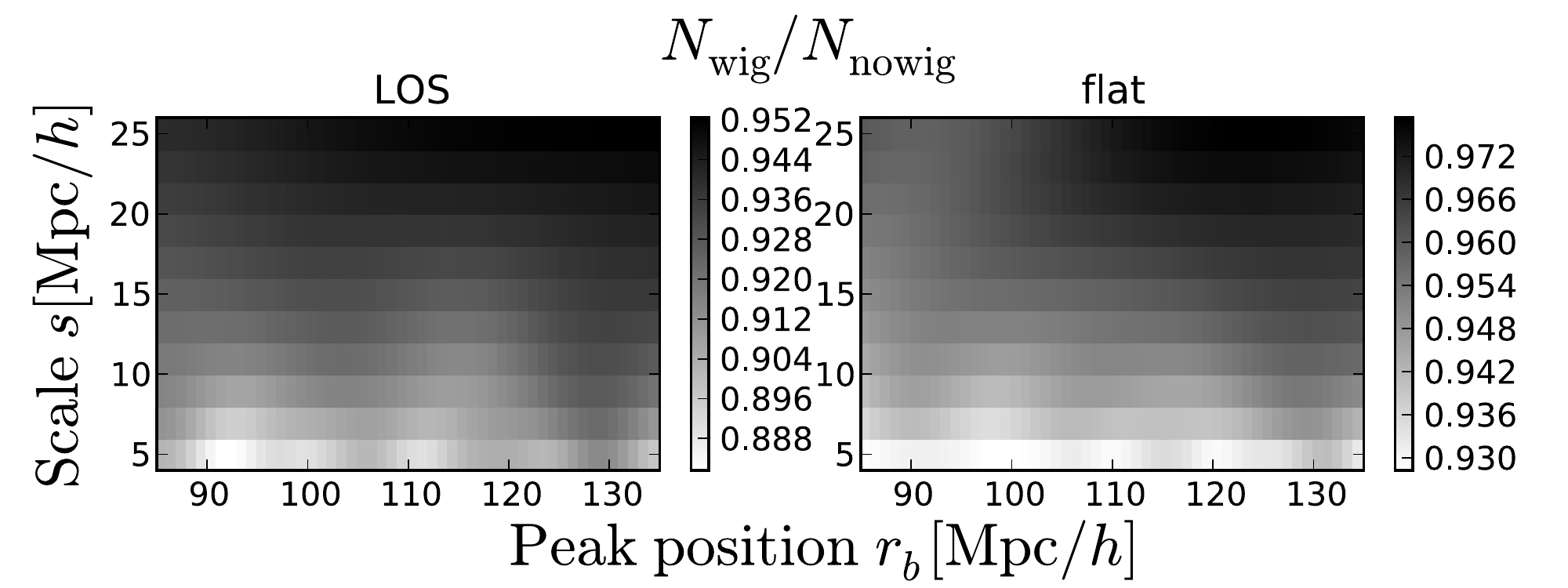}
  \end{center}
\caption{The ratio of the noise $N$, i.e.\ the standard deviation of
    the wavelet as a function of $(r_b,s)$, for Gaussian simulations
    with and without a BAO feature.
    \label{fig:wignoiserat}
  }
\end{figure}

A proper test against the null hypothesis of no BAO feature would use
the variance from a no-wiggle power spectrum, but instead, below we
estimate variances from the sample itself.  Thus it is relevant to ask
how the BAO feature affects the standard deviation of the wavelet
transform.  Fig.\ \ref{fig:wignoiserat} shows the ratio of the noise
with and without the BAO feature, for both LOS and flat weighting.
Typically, the ratio is within 10\% of unity, with only very slight
fluctuations in $r_b$.

\subsection{Peak-finding Statistics and Systematics in Gaussian fields}

In each simulation, we find the maximum of its wavelet transform over
a grid $(r_b\in [85,135], s\in [2,26])$, with a grid spacing of
$2\hMpc$.  We require that peaks not be along a rectangle-edge in $s$
or $r_b$; a peak along an edge might indicate that a true peak would
be outside the area searched.  Thus, the actual possible range for
peaks is a bit smaller, $(r_b\in [87,133], s\in [4,24])$.  In Bayesian
language, this is the range of our flat prior.  The error bars that we
quote in $r_b$ are marginalized over this flat prior in $s$.

One might worry about covariance as a function of in the S/N plot, and whether it
affects our estimate of either the significance, or location, of the
bump.  Certainly, the wavelet coefficients at nearby $(r_b,s)$ are
correlated (though less-so than in the raw $\xi$).  The smoothness of
the S/N plots is a sign of this.  But this correlation does not affect
the interpretation of the S/N: at $(r_b,s)$, it is a good measure of
the confidence that a bump exists there, specifically that the Mexican
hat wavelet coefficient is positive at that $(r_b,s)$.  It does not
matter that nearby in $r_b$ and $s$, the S/N is likely quite similar.
Covariance in the wavelet coefficient also does not affect the peak
location estimate, except that a particular wavelet choice could
produce a broader-than-optimal peak in the S/N plot.  We have not
attempted to optimize or orthogonalize our wavelet, but as we find
below, the statistical and systematic errors on the true bump location
in Gaussian simulations are still small.

\subsubsection{Statistical errors}
\begin{figure}
  \begin{center}
  \includegraphics[scale=0.44]{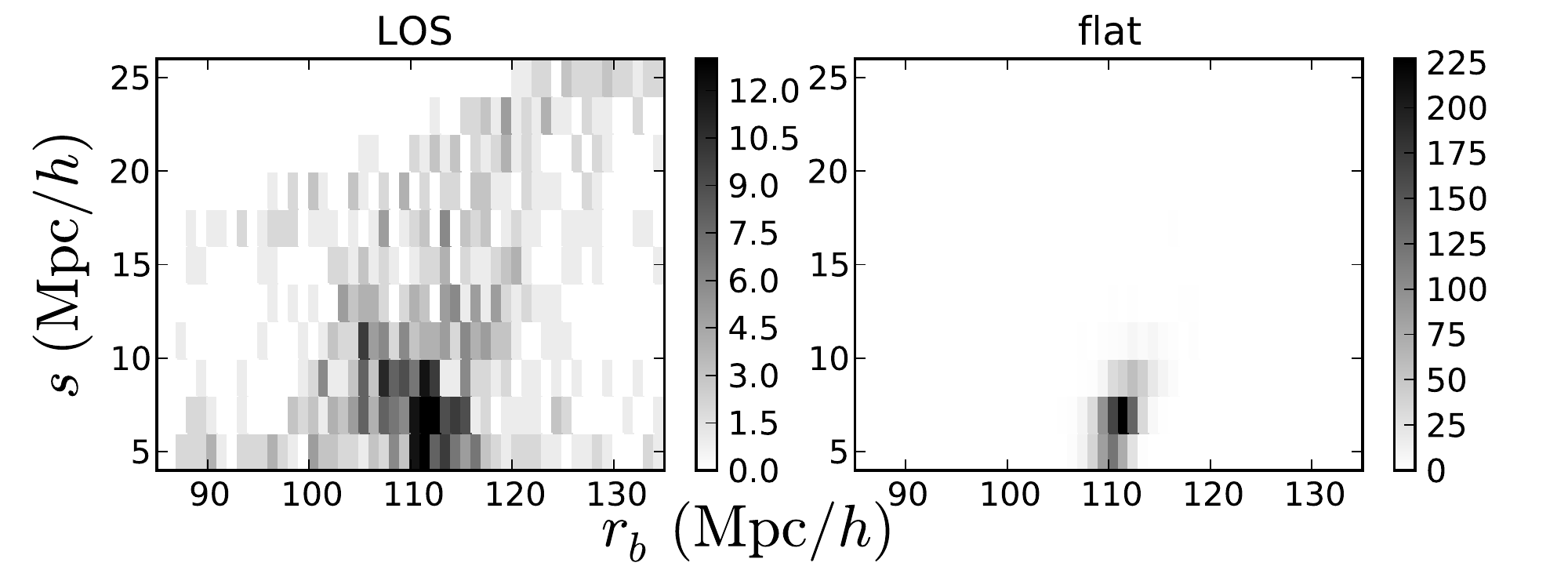}
  \end{center}
\caption{A 2D histogram of the peak in $(r_b,s)$ of
    $\avg{G_2(r_b,s)\xi_i(r,\theta)}$ measured in 1400 Gaussian
    simulations $i$, divided by its simulation-to-simulation standard
    deviation.  Here $\xi$ is the full 3D correlation function, with
    linear redshift distortions applied.  With LOS and flat
    weightings, the peaks have locations 111.2$\pm 9.0$ and
    110.7$\pm 1.6\hMpc$.
    \label{fig:peaks}
  }
\end{figure}

\begin{figure}
  \begin{center}
  \includegraphics[scale=0.44]{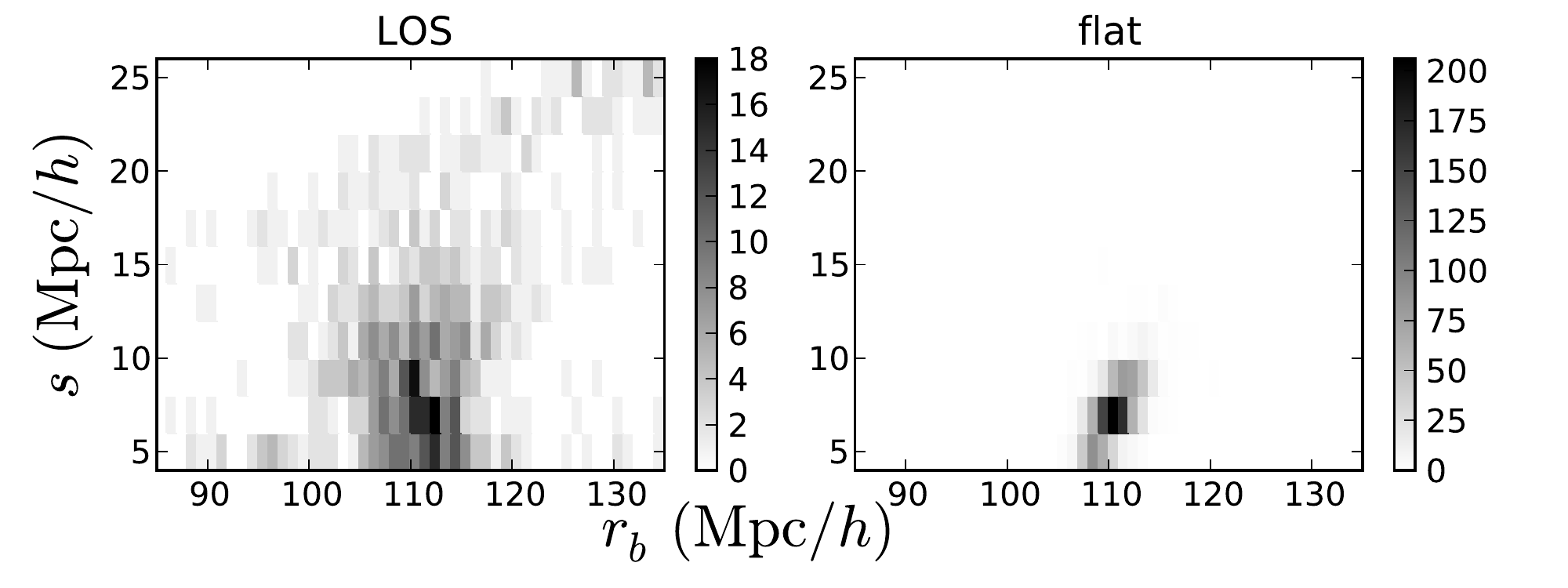}
  \end{center}
\caption{Same as Fig.\ \ref{fig:peaks}, for averaged 2D slice
    correlation functions.  The S/N assumes that a full box volume is
    being analyzed.  With LOS and flat weightings, the peaks have
    locations 111.1$\pm 7.9$ and 110.2$\pm 1.8\hMpc$.
    \label{fig:peaks_slice16}
  }
\end{figure}

\begin{figure}
  \begin{center}
  \includegraphics[scale=0.44]{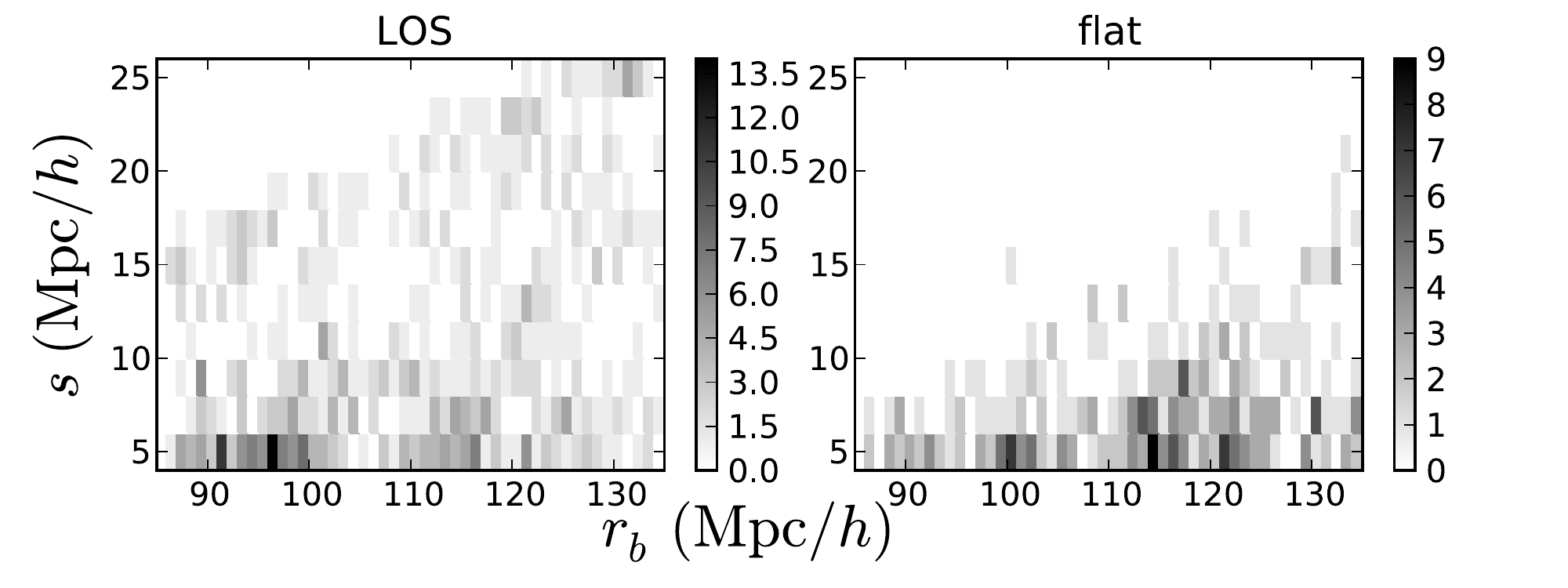}
  \end{center}
\caption{Same as Fig.\ \ref{fig:peaks}, except where the Gaussian
    simulations are generated using a no-wiggle power spectrum.  With
    LOS and flat weightings, the peaks have locations 109.3$\pm
    14.1$ and 113.8$\pm 12.2\hMpc$.
    \label{fig:peaks_ehu}
  }
\end{figure}

\begin{figure}
  \begin{center}
  \includegraphics[scale=0.44]{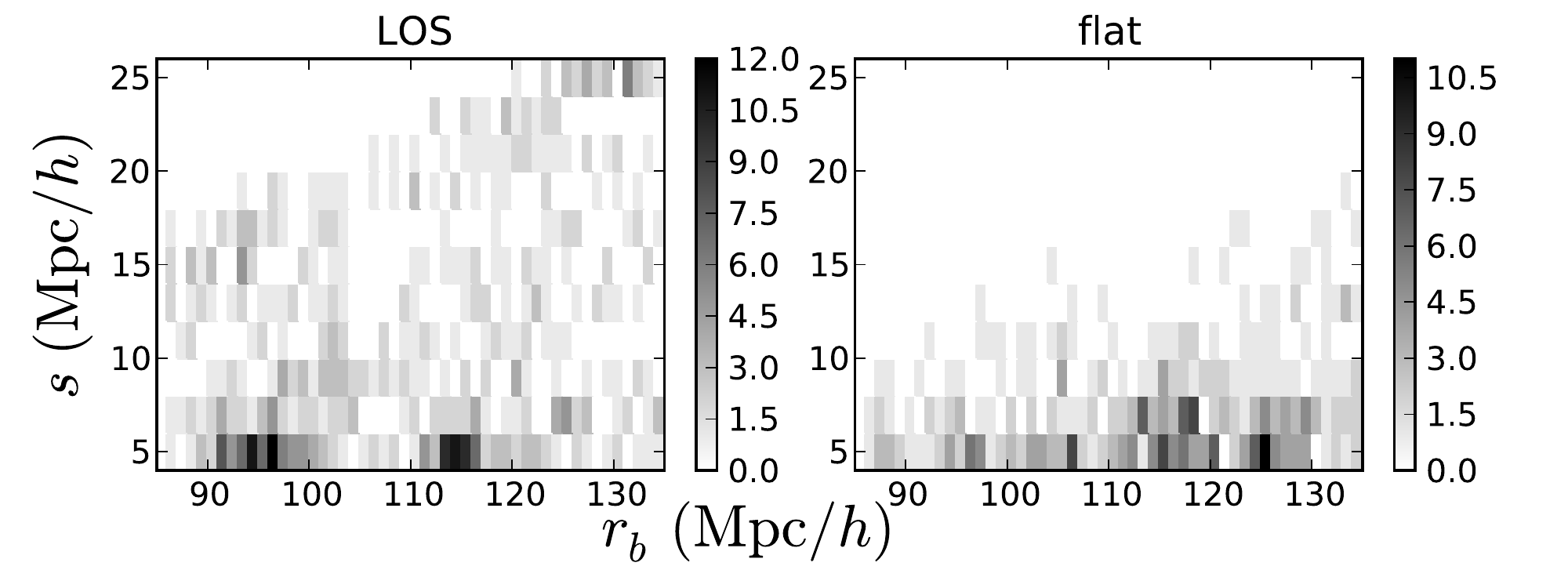}
  \end{center}
\caption[1]{Same as Fig.\ \ref{fig:peaks}, for averaged 2D slice
    correlation functions, and where the Gaussian simulations are
    generated using a no-wiggle power spectrum.  With LOS and flat
    weightings, the peaks have locations 109.2$\pm 13.6$ and 114.1$\pm
    12.6\hMpc$.
    \label{fig:peaks_ehu_slice16}
  }
\end{figure}

Figs.\ \ref{fig:peaks}, \ref{fig:peaks_slice16}, \ref{fig:peaks_ehu}
and \ref{fig:peaks_ehu_slice16} show 2D histograms of peak positions
for individual redshift-distorted realizations in the ensembles whose
mean S/N plots appear on the right-hand sides of
Figs.\ \ref{fig:gauss}, \ref{fig:gauss_slice16}, \ref{fig:gauss_ehu}
and \ref{fig:gauss_ehu_slice16}.  The case most relevant to our SDSS
measurements is shown in Fig.\ \ref{fig:peaks_slice16}, in which
averaged 2D slice correlation functions, drawn from a simulations with
BAO features, are analyzed.

In the presence of a BAO feature, the (posterior) distributions of
peak $r_b$'s after peak-finding are tightened considerably compared to
the prior.  The set of $r_b$'s in the flat prior have a standard
deviation of 14$\hMpc$, which contracts to 8$\hMpc$ when looking along
the LOS, and to 2$\hMpc$ (we conservatively round up) in the case of
flat weighting.  In contrast, if there is no BAO feature, the
posterior distribution of $r_b$ is hardly narrower than the prior.  In
the left panel of Fig.\ \ref{fig:peaks_slice16}, there are a couple of
curious accumulations of very narrow peaks, along the $s=4$ edge.
However, they do not appear to be related to the peaks in the cases
with BAO features.  True features are somewhat extended in $s$; we
suspect that these narrow features are simply noise.

As recently stated by \citet{cabre10}, the model-comparison question
of whether a BAO feature exists in a given sample is quite different
than the question of its constraining power.  Under the reasonable
assumption that the power spectrum underlying the structure in our
Universe has a BAO feature, even a low-significance bump gives
substantial constraining power.  For example, if there happens to be a
LOS wavelet S/N peak in a sample, we find that the LOS weighting gives
an $8\hMpc$ error bar, even though about half of the mocks possess no
LOS peak at all (see Fig.\ \ref{fig:peakheights}).


One might also wonder whether error ellipses on the peak $r_b$ and $s$
can be inferred directly from the S/N plots.  Comparing the S/N plots
to the distributions of peaks in Figs.\ \ref{fig:peaks},
\ref{fig:peaks_slice16}, \ref{fig:peaks_ehu} and
\ref{fig:peaks_ehu_slice16}, it seems that there is only a loose
relationship.  For the two to correspond exactly, the probability of a
given $(r_b,s)$ having a global peak would have to be simply related
to its local S/N value.  Given the additional complication of
correlations in the S/N plot, it is perhaps not surprising that they
are not simply related.  One curious difference is that peaks seem
more prevalent at small $s$ than the mean S/N plots would suggest,
especially for flat angular weighting.

\subsubsection{Frequency of maximum S/N}
\label{sec:freq_max}
The previous section concerned the expectation value of the S/N in the
wavelet transform.  But if the peak in S/N is broad, the maximum S/N
measured in an individual realization will tend to exceed the maximum
in the mean, since nearby coefficients may fluctuate independently,
and each might produce the observed peak.

\begin{figure}
  \begin{center}
  \includegraphics[scale=0.42]{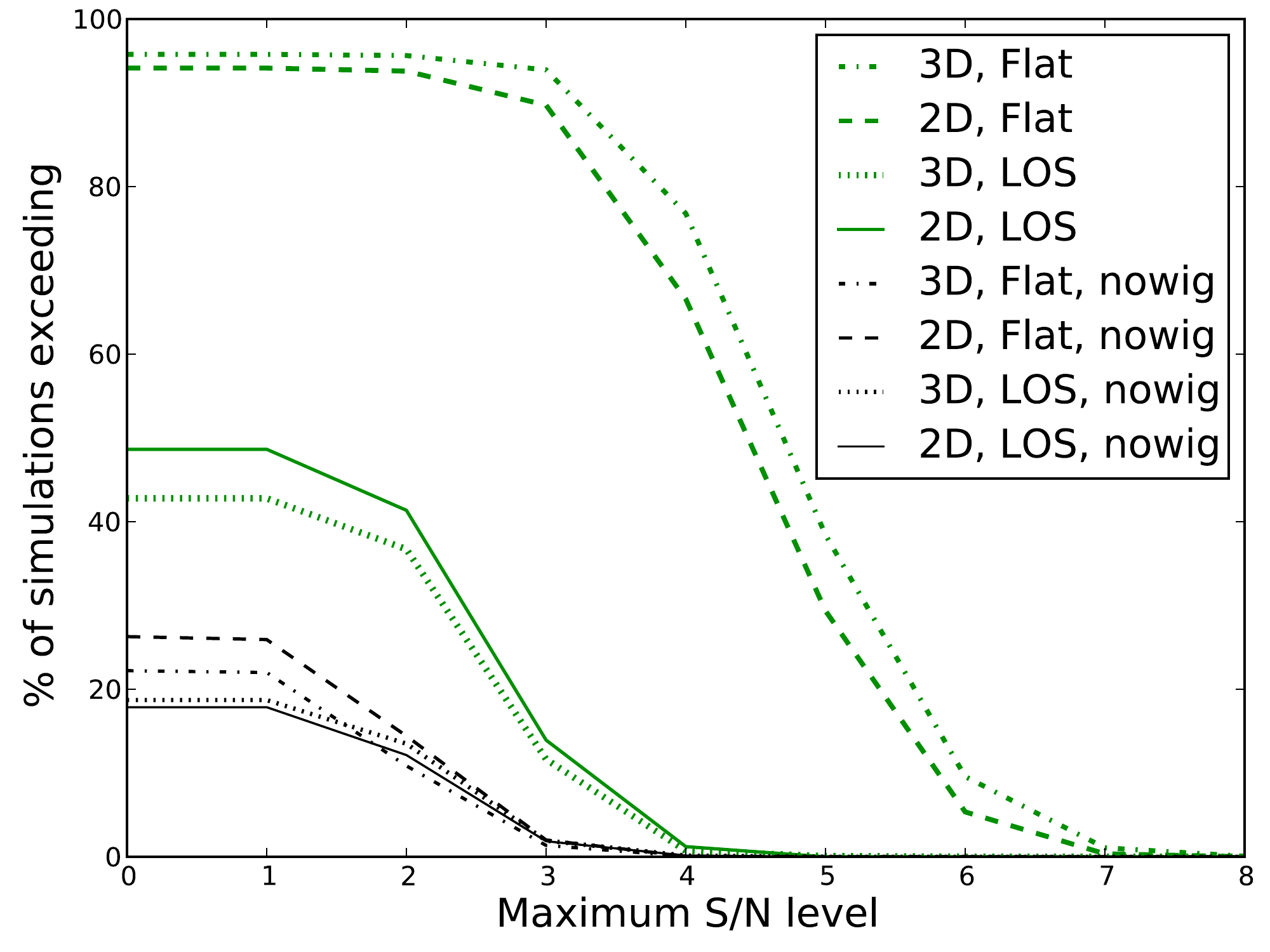}
  \end{center}
\caption[1]{Cumulative frequencies of maximum wavelet S/N in
    Gaussian simulations of approximately the volume of the SDSS
    sample, in various cases.  For the green curves, simulations were
    generated using a \camb\ power spectrum, and for the black, a
    no-wiggle power spectrum was used.
    \label{fig:peakheights}
  }
\end{figure}

Fig.\ \ref{fig:peakheights} shows the cumulative frequency over 1400
Gaussian simulations of the maximum S/N in various scenarios.  For
this plot, peaks in $r_b$ were sought in a narrower range of $r_b$
than in the previous sections, more indicative of the peaks actually
observed in the SDSS sample.  As previously, peaks were excluded if
they were along an edge in $s$ or $r_b$; the actual range of peaks
allowed was 100 to 120$\hMpc$.  The fraction of simulations with a
true peak (not on an edge of the prior) can be read off as the
fraction with ${\rm S/N}= 0$.

From this figure, we see that if the underlying power spectrum has a
BAO feature, using our method it is not uncommon to get a peak up to 3
$\sigma$ from a Gaussian sample with the SDSS volume, even along the
LOS.  Below we estimate the S/N of our bump detection at the peak
($r_b,s$) to be 2.2 (LOS weighting) and 4.0 (flat weighting).  If the
underlying power spectrum has a BAO feature, about 40\% of the
simulations had a LOS peak of \mbox{S/N $\ge 2.0$}, and about 70\% had
a peak with flat weighting of \mbox{S/N $\ge 4.0$}.

On the other hand, if the underlying power spectrum does not have a
BAO feature (i.e.\ in the null hypothesis of BAO peak detection), it
becomes much more uncommon to get such peaks.  About 11\% of no-wiggle
simulations had a LOS peak of S/N $\ge 2.2$, and about 0.2\% of
no-wiggle simulations had a peak with flat weighting of S/N $\ge 4.0$.

\subsubsection{Systematic errors}

\begin{figure}
  \begin{center}
  \includegraphics[scale=0.3]{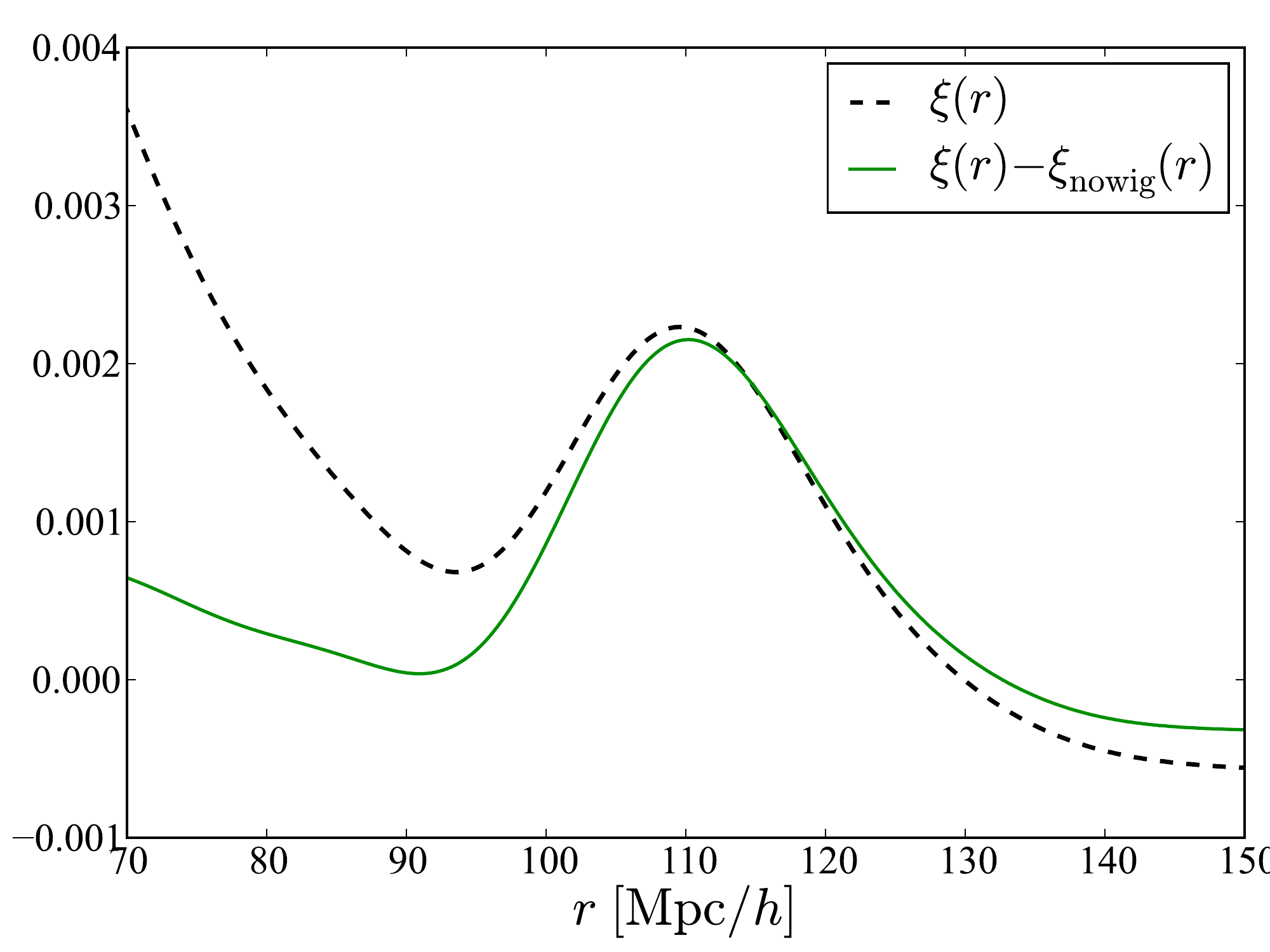}
  \end{center}
\caption[1]{A linear $\xi(r)$ from \camb\ (with peak at
    $109.5\hMpc$), along with the difference between the \camb\ and
    the no-wiggle linear correlation functions (with peak at
    $110.3\hMpc$).
    \label{fig:xidiff}
  }
\end{figure}

The Mexican-hat wavelet transform is a natural measure of the second
derivative at position $r_b$, if the function it is applied to is
smoothed over a scale $s$.  In fact, arguably a peak in the
Mexican-hat wavelet coefficient would be a good definition of the peak
position itself, if at an appropriate scale.  An alternative
definition is the actual local maximum of the peak in the linear
correlation function.  However, this definition can cause a bias, if
the peak is on top of a slope, as mildly occurs in the case of a
\lcdm\ correlation function.  Fig.\ \ref{fig:xidiff} shows $\xi_{\rm
  CAMB}(r)$ computed using a \camb\ power spectrum, along with the
difference between that and a no-wiggle $\xi_{\rm nowig}(r)$.  Both
are measured from density fields with exactly the Fourier amplitudes
prescribed by their linear power spectra.  The peaks of $\xi_{\rm
  CAMB}(r)$, and of the difference, are at 109.5 and 110.3$\hMpc$, a
shift of almost 1$\hMpc$.

The means in the peak distributions shown in Figs.\ \ref{fig:gauss}
and \ref{fig:gauss_slice16} are within $\sim1\hMpc$ of both peaks
shown in Fig.\ \ref{fig:xidiff}, if the peak is defined as the actual
local maximum of the peak in $\xi$, the wavelet estimator tends to
overestimate the peak location by about $1\hMpc$.

Again, the small systematic errors found here come from purely
Gaussian simulations, without nonlinearities or galaxy bias, which
could generate their own systematic errors, perhaps of order the
statistical error bars in our sample.

\section{Measurements from the Millennium simulation}

\begin{figure*}
  \begin{center}
    \includegraphics[scale=0.6]{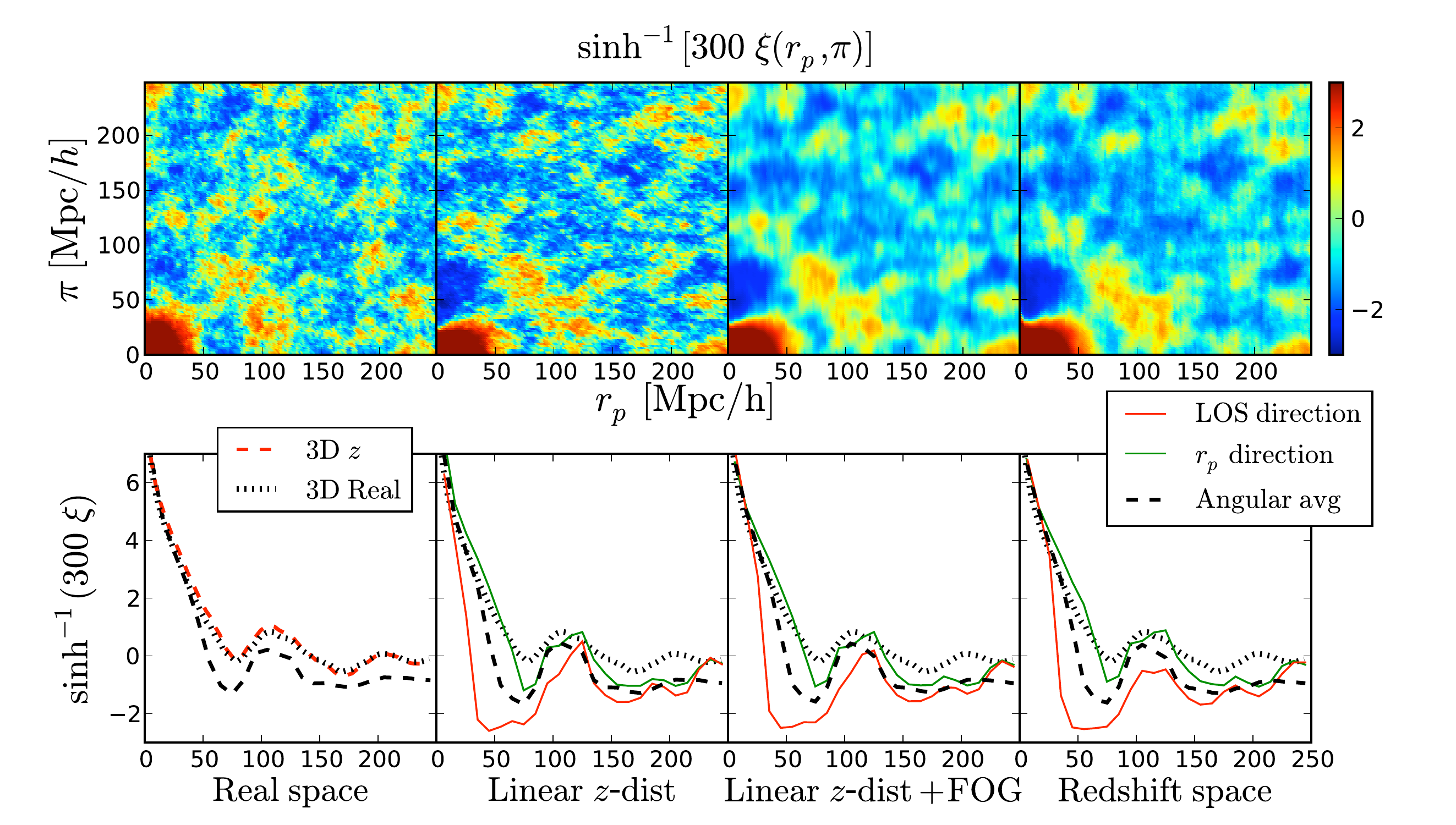}
  \end{center}
\caption{Correlation functions of a sample of galaxies modeled
    within the Millennium simulation (MS).  For reference, the 3D
    real-space correlation function is shown in dotted black in all
    lower panels.  The colored solid curves in the bottom panels show
    the angular dependence of $\xi(\pi,r_p)$, shown with its full 2D
    structure in the top panels.  The 2 angular bins include angles
    within $6\degr$ of the LOS (red) and of the direction
    perpendicular to it (green).}\vspace{0.1in}
  \label{fig:millpisigma}
\end{figure*}

We compare our results from the observations with simulations of
large-scale structure formation.  We used Millennium-simulation (MS)
galaxies to investigate realistic redshift distortions of galaxy
samples.  However, its box size ($500\hMpc$), smaller than the size of
our sample, and the fact that we only have a single realization, make
it insufficient to derive strong conclusions about the detailed
statistics of BAO measurements.  So, we also employed Gaussian
simulations to get a better idea of the correlations involved, keeping
in mind that non-linearities would likely degrade significances and
constraints we measure.  

Figure \ref{fig:millpisigma} shows various correlation functions
of galaxies from the MS, measured on a $256^3$ grid.  The galaxies
used were brighter than an $r$-magnitude of $-20$ (absolute) as modeled
by \citet{delucia}, giving a mean density of 0.02 $(\hMpcnosp)^{-3}$.

The curves shown are averages of slice correlation functions $\xi$
over all possible orientations, axes, and slice locations.  For the
real-space $\xi$'s, the grid was split into 64 slices of thickness
$7.8125\hMpc$ along all three Cartesian axes, giving $3\times
64\ \xi$'s to average together.  In the middle two columns,
linear redshift-space distortions were generated analytically, using
$(\beta=0.46,\sigma=0)$ and $(\beta=0.46,\sigma=3\hMpc)$ in the
Kaiser formula 
\begin{equation}
  \delta_k^{z}=\delta_k \frac{1+\beta \mu^2}{1+(k\sigma\mu)^2},
  \label{eqn:linearz}
\end{equation}
where $\mu=\cos \theta$. In the right column, redshift-space
distortions were generated using the actual MS galaxy velocities, not
an approximate model.  In each case, the result is an average over all
possible permutations among the three axes of the LOS and the two axes
along which the slices were cut.

Generally, linear-theory redshift-space distortions in 2D slices do
appear to sharpen the baryon bump in $\xi(r)$ relative to real space.
Indeed, as we find below with Gaussian simulations, linear
redshift-space distortions seem not only to sharpen the peak, but to
increase its robustness too, at least when it is analyzed using
wavelets.  At some level, one expects fingers of God (FoG) tend to
degrade all features in $\xi(\pi,r_p)$, even away from the LOS.
Interestingly, though, at least in the MS, the 1D curves do not change
much between the two bottom-middle plots, suggesting that the
fogginess effected by FoG may be minor.

\begin{figure}
  \begin{center} \includegraphics[scale=0.44]{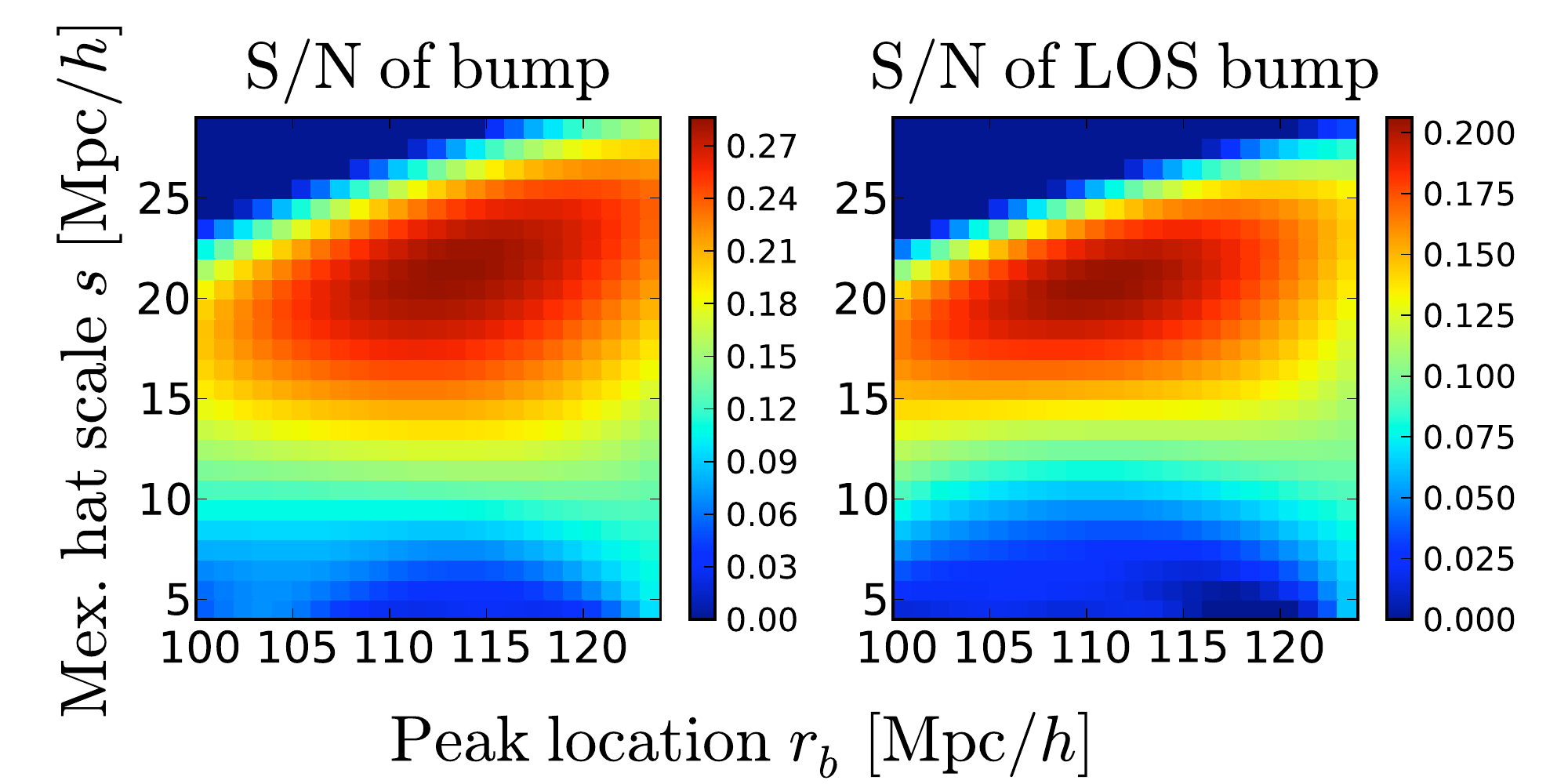}
  \end{center}
\caption{The signal-to-noise ratio of the wavelet transform
    $\avg{G_2(r_b,s)\xi(\pi,r_p)}$ in the MS simulation.  In the MS,
    the variances are measured among slices.  The left column uses a
    flat weighting in $\theta$, while the right column includes only
    $\theta<6\degr$.  With flat and LOS weightings, the peaks are at
    $(r_b,s)=(110, 20)$ and (113, 21), respectively.
    \label{fig:mill_numsigma}
  }
\end{figure}

We applied the wavelet analysis to the redshift-space MS simulations,
as well.  Fig.\ \ref{fig:mill_numsigma} shows the mean S/N from these
samples.  With flat weighting in angle, there is a $3\hMpc$ shift
relative to the Gaussian peak location.  It would be tempting to use
this as an estimate of the systematic error from using a simulation
with full non-linearities and galaxy formation, but it could very well
be a statistical fluctuation instead.

\begin{figure}
  \begin{center}
  \includegraphics[scale=0.4]{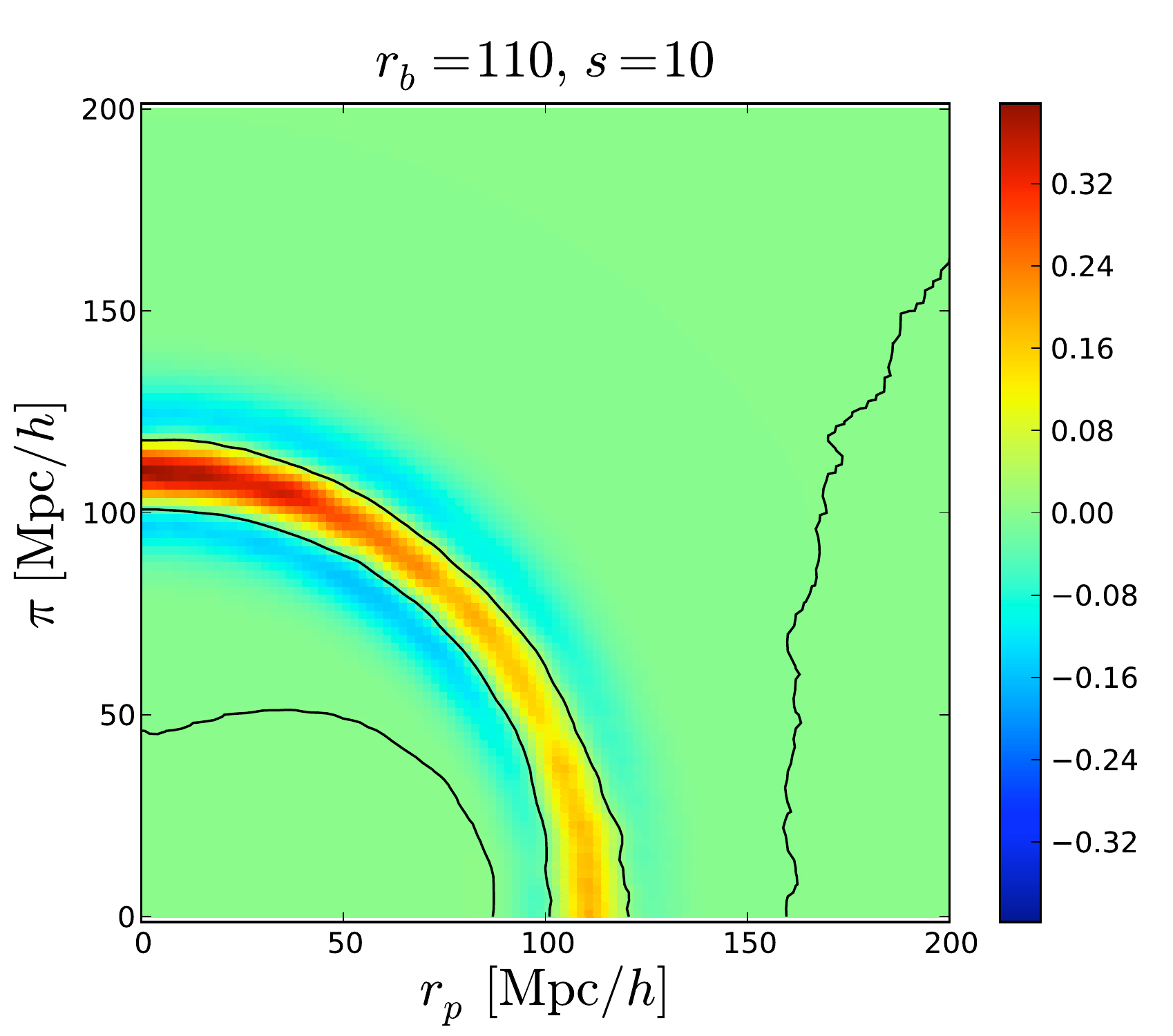}
  \includegraphics[scale=0.4]{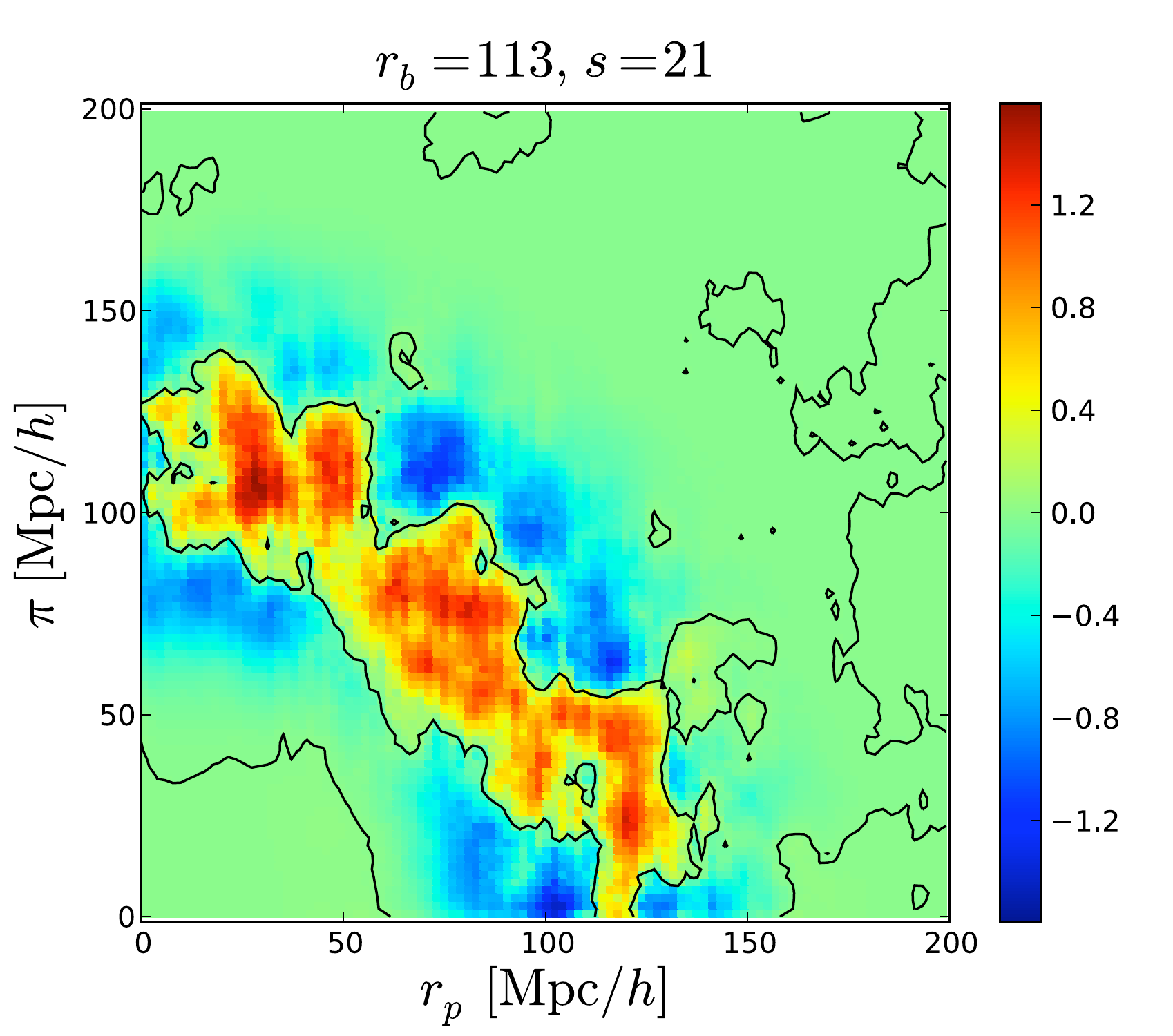}
  \end{center}
\caption{The top panel shows the linear-theory redshift-space
    correlation function, flattened at $(r_b,s)=(110, 10)\hMpc$, as
    measured from 2D slices from Gaussian simulations discussed in
    Sect.\ \ref{sec:wavgauss}.  The bottom panel shows the MS galaxy
    redshift-space correlation function flattened at
    $(r_b,s)=(113,21)\hMpc$.  They are flattened at the same $r_b,s$
    as their respective peaks in wavelet S/N, measured with flat
    weighting.  For comparison with other figures, they show
    $\sinh^{-1}(300\xi)$ instead of $\xi$, although here $\xi$ is
    small, so the $\sinh^{-1}$ transform hardly changes the plot's
    appearance.
 \label{fig:flattening}
 }
\end{figure}

Fig.\ \ref{fig:flattening} shows measurements of
$G_0(r_b,s)\bar\xi(r_b,s;\pi,r_p)$ for MS galaxies, and in linear
theory.  In these figures, the values of $r_b$ and $s$ are those at
the peaks in the signal-to-noise ratio of the wavelet transform
$\avg{G_2(r_b,s)\xi}$.

\section{Measurements from the SDSS}

\subsection{The Sample}

We analyzed the SDSS DR7 \citep{dr7} main-galaxy sample
\citep[MGS][]{mgs}.  These are all galaxies observed by SDSS that have
an R-band Petrosian magnitude R$\le 17.77$.  Importantly for our
analysis, their redshifts have been measured spectroscopically.  We
selected galaxies designated as sciencePrimary, from the Northern cap
of SDSS, in stripes 9 through 37, with a redshift confidence $>$0.9,
and redshift error $<$0.1.  We also cut galaxies from the tails of the
selection function, including only galaxies with distance
$100\hMpc<r<750\hMpc$.  This gave us a total of 527,781 galaxies to
begin with. Furthermore, there were several regions with 'holes' and
small, incompletely sampled regions in stripes 13, 29, 35 and 36. We
removed all objects in these incomplete areas.  This left us with a
total of 527,362 objects.

We have computed comoving radial distances $r$ for each object,
expressed in $\hMpc$, using built-in functions in the SDSS SkyServer
database \citep{manu10}, with $\Omega_m=0.279, \Omega_\Lambda=0.721$,
and $w_0=-1$.  The angular coordinates were converted to Cartesian,
and rotated to a coordinate system, whose z-axis was along the SDSS
North Pole, at ra=185, dec=32.5. Each of the normal vectors were then
multiplied with the computed radial distance to give us a 3D location.
 
\begin{figure}
  \begin{center}
  \includegraphics[scale=0.4]{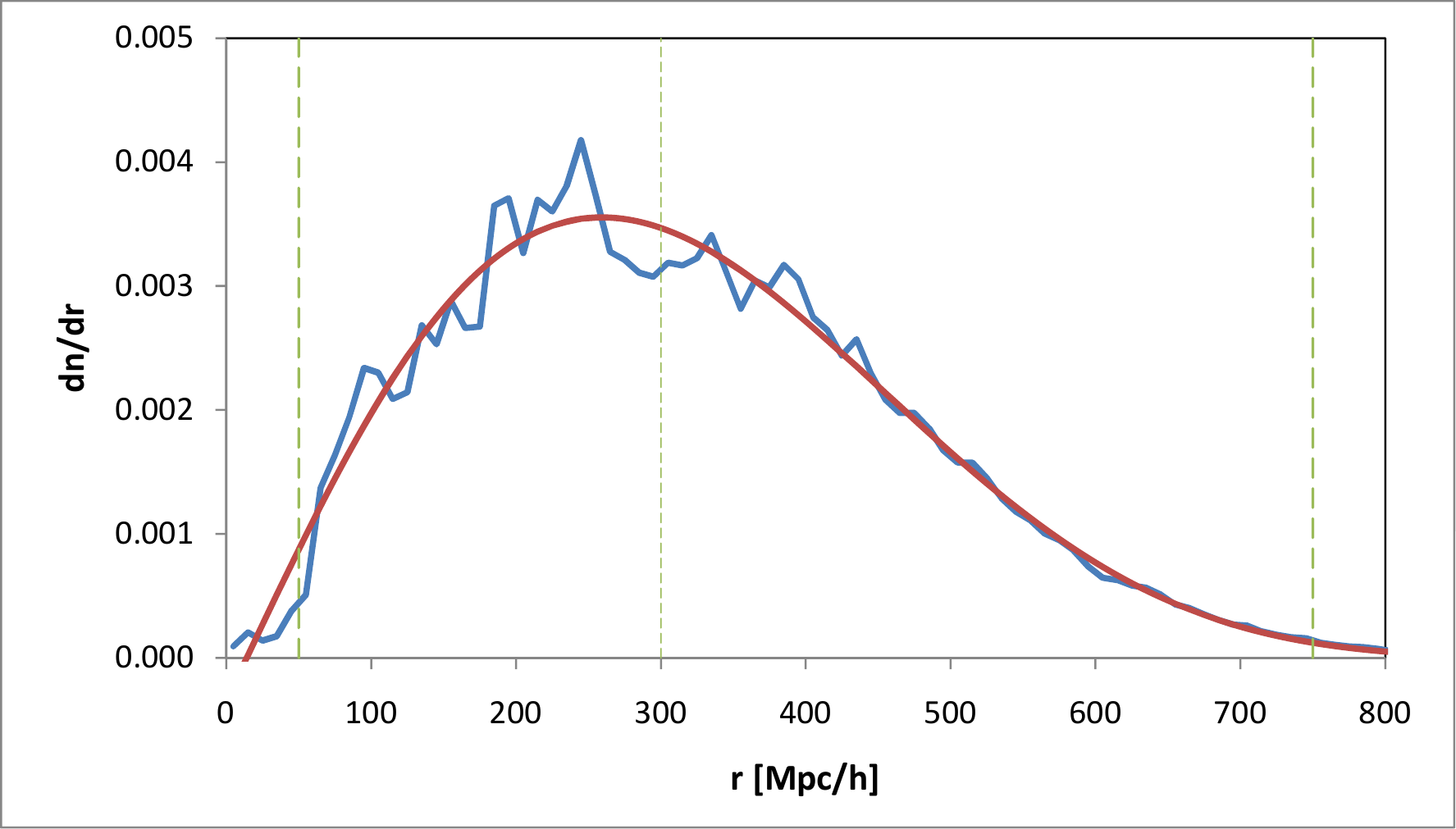}
  \end{center}
\caption{The radial distribution function $dn(r)/dr$ of the MGS
    sample we used, normalized to $\int n(r) dr=1$ (blue), and a
    polynomial fit (red).  The polynomial fit was used to generate our
    random samples.  We also show the near and far cuts for the full
    sample, and the near cut for the high-$z$ sample.  The large
    upward fluctuation at about 220$\hMpc$ is produced by the Sloan
    Great Wall.
 \label{fig:selfun}
  }
\end{figure}

We computed a smooth polynomial fit to the redshift distribution,
$dn(r)/dr$, using a 6th order polynomial in $r$. The curve is shown in
Fig.\ \ref{fig:selfun}. We used the angular selection mask, defined by
boundaries of the selected stripes and the censored 'holes' and this
analytic radial distribution to generate 17M random galaxies with the
correct geometric properties. Each random galaxy had an additional
random number precomputed, making it easy to select decimated subsets
for further analyses.

These two data sets were then stored in two database tables. We
created a simple function that was able to take an arbitrary rotation
of the samples around the $z$-axis. We stepped the rotation angle in
15$\degr$ increments, from $0\degr$ to $165\degr$, for a total of 12
angular orientations. For each orientation we extract $2.5\degr$ thick
slices, resembling the original SDSS stripes, except for the
rotation. We eliminated slices which contained only a small fraction
of the data, located at the edges of the survey, i.e.\ slices of width
$<20\degr$.  Slices that were wider than $80\degr$ were split in half
across, so that no slice exceeded the width of $80\degr$. This gave us
661 slices.

\subsection{Computing the Correlations}

We used the \citet{ls93} estimator to estimate the correlations, for
its optimal behavior \citep{kerscher00}:
\begin{equation}
  \hat\xi_{LS}=\frac{DD-2DR+RR}{RR},
\end{equation}
where $DD$, $DR$ and $RR$ are numbers of pairs of random and actual (data)
galaxies, in length bins of $\xi$.  

Measuring the distances among all random and data galaxies is
inherently an $N^2$ problem. The current state-of-the-art solution by
\citet{moore01} involves binary trees built on the datasets, and uses
a dual-tree traversal algorithm. The idea is to speed up the procedure
by checking distance constraints on pairs of tree-nodes that represent
3-D boxes. If all pairs of points coming from the cells fit in a
single bin, one can increment the counts and stop going deeper on
that branch.
For low-resolution 1-D statistics, such as the angular correlation
functions measured in a dozen logarithmic bins, the above procedure
can indeed increase the performance tremendously especially when one
is interested in small-scale clustering and able to discard early all
pairs of large separations.
Our problem is more difficult: here we measure the the two-dimensional
redshift-space $\pi$-$r_p$ correlation function out to the largest
scales at high resolution in $800 \times 800 = 640$ 000 bins of
spacing $0.5$ Mpc. (To eliminate null pixels in the region of
interest, however, we ended up degrading the resolution to 2 Mpc.)
The dual-tree code slows down in this high-bin limit and becomes
essentially as expensive as the brute-force naive method.

Our solution is to implement the counting on modern graphics processing
units (GPUs) that offer hundreds of cores and run tens of thousands of
threads simultaneously on commercial video cards.
We use NVIDIA's Compute Unified Device Architecture (CUDA) to
implement the parallel correlation function code in the C++
programming language and integrate it with SQL Server where the data
reside.  Using SQL wrapper routines, we run the analysis directly on
GTX 295 GPUs without temporary intermediate file storage.  The
performance is hundreds of times faster on this parallel architecture
when compared to todays CPUs, which is not too surprising for the
large number of algorithm logic units (480 ALUs) on these cards.  The
results from the GPUs are returned in database tables, stored and
further analyzed in SQL to compute the final correlation functions.

The 2D correlation function was computed for each slice, with a
2$\hMpc$ resolution, out to 570$\hMpc$ in each direction. A total of
400 trillion galaxy pairs were computed, including both real and
random points.

\subsection{Results from the SDSS}

\begin{figure}
  \begin{center}
    \includegraphics[scale=0.42]{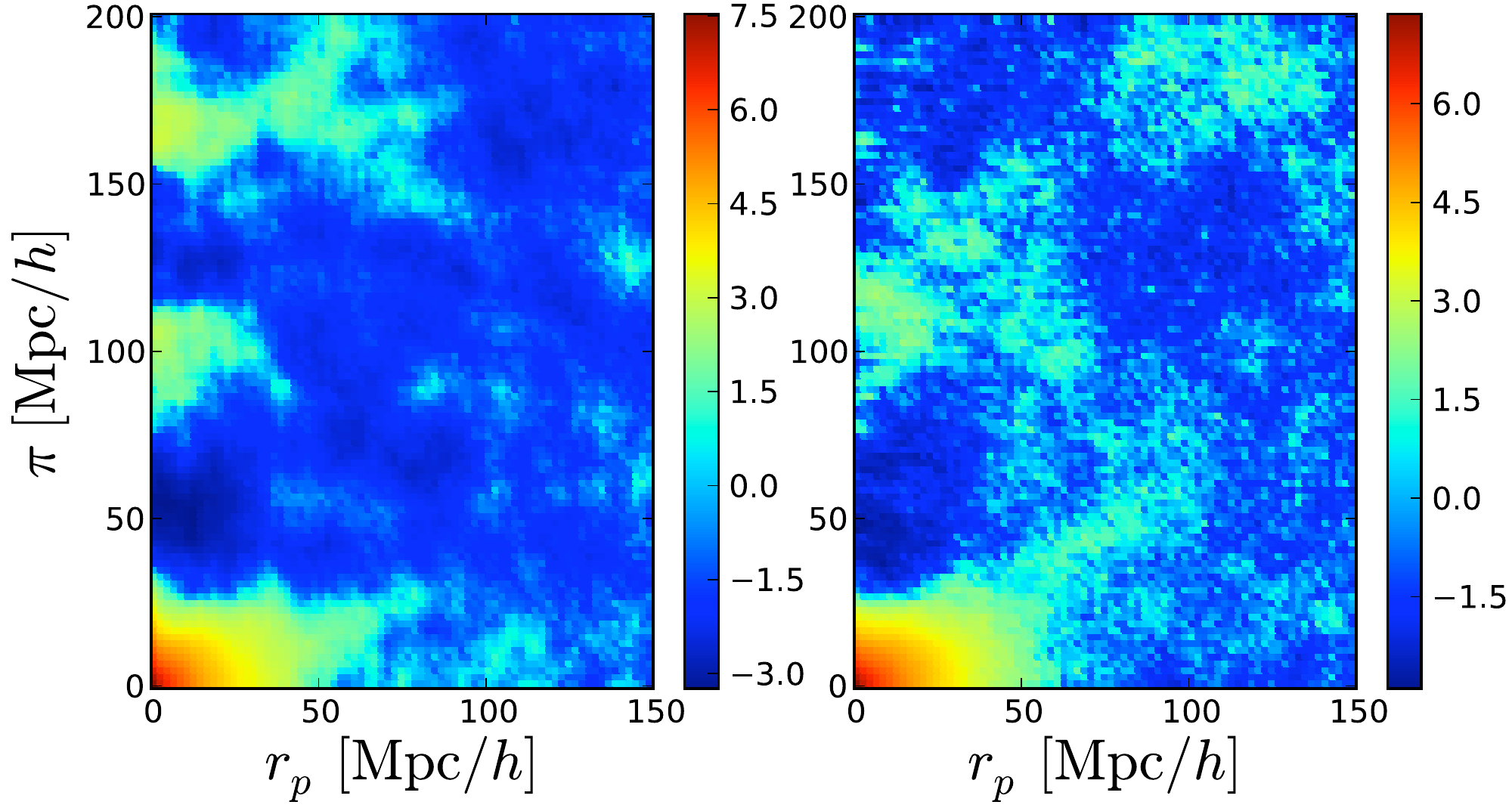}
  \end{center}
\caption{The correlation function $\xi(\pi,r_p)$ measured from the
    SDSS main-galaxy sample.  The left panel is measured from the
    $100-750\hMpc$ full sample, and the right panel is from the
    $300-750\hMpc$ high-$z$ sample.  The high-$z$ sample is only a bit
    smaller (losing 6\% by volume), but it excludes the Sloan Great
    Wall, a structure whose inclusion is known to alter clustering
    statistics substantially.  The high-$z$ sample appears grainier
    than the full sample because of increased shot noise.}
 \label{fig:xi2d_old_new}
\end{figure}

Fig.\ \ref{fig:xi2d_old_new} shows $\xi(\pi,r_p)$ from two samples:
first, the full sample, including all galaxies from $100-750\hMpc$,
and a high-redshift sample, $300-750\hMpc$.  Along the LOS, the full
sample has two prominent peaks: one at about $97\hMpc$, and a second
at about $170\hMpc$.  The first could be associated with a BAO
feature, but the second could not, given plausible priors on the BAO
scale from previous measurements \citep[e.g.\ ][]{e05}.

We analyzed the higher-redshift sample because the measurement gives
perhaps undue weight to the densely sampled structures at low
redshift, such as the Sloan Great Wall.  Including the Sloan Great
Wall, which is at a distance of about 220$\hMpc$ \citep{gott05}, is
known to change clustering statistics substantially
\citep[e.g.\ ][]{nichol06}, producing an upward fluctuation in
Fig.\ \ref{fig:selfun} of 20\%.

A better solution might be to reduce the additional cosmic variance
from such a structure by down-weighting it, e.g.\ with a Gaussianizing
density mapping \citep{nss09}, but for the present paper we simply cut
the near part of the sample, within 300$\hMpc$, and apply an
additional weighting to each galaxy. By volume, this high-$z$ sample
is only 6\% smaller than the the full one, although it only contains
about half the galaxies, 261,737.  The effective volume including shot
noise \citep{tegmark97}
\begin{equation}
  V_{\rm eff}(k)=\int\left\{1+\left[\frac{dn(r)}{dr}P(k)\right]^{-1}\right\}^{-2}d^3{\bf r}
\end{equation}
for the full sample is $0.29\ (\hGpcnosp)^3$.  For the high-$z$
sample, $V_{\rm eff}=0.27\ (\hGpcnosp)^3$, a 9\% difference.  Here we
use $P(k=0.1\ihMpc)= 7000\ (\hMpcnosp)^3$ (from the linear power
spectrum used in the Gaussian simulations, at about the BAO scale in
wavenumber), assuming a $\pi$-steradian survey.

The optimum (minimum variance) spatial weighting for galaxies at
radial distance $r$, on the clustering scale $L$ is given by
\citep{k86}
\begin{equation}
w(r,L) = \frac{1}{1+4\pi {\bar n(r)} J_3(L)}.
\end{equation}
For the SDSS main-galaxy sample, $J_3(110)\approx 30,000$. Using the
selection function of the galaxies we can compute the weight
corresponding to the 110$\hMpc$ scale. We then fit a third-order
polynomial to the log of the weight function, and use this analytic
expression in the further analysis.

\begin{figure}
 \begin{center}
   \includegraphics[scale=0.3]{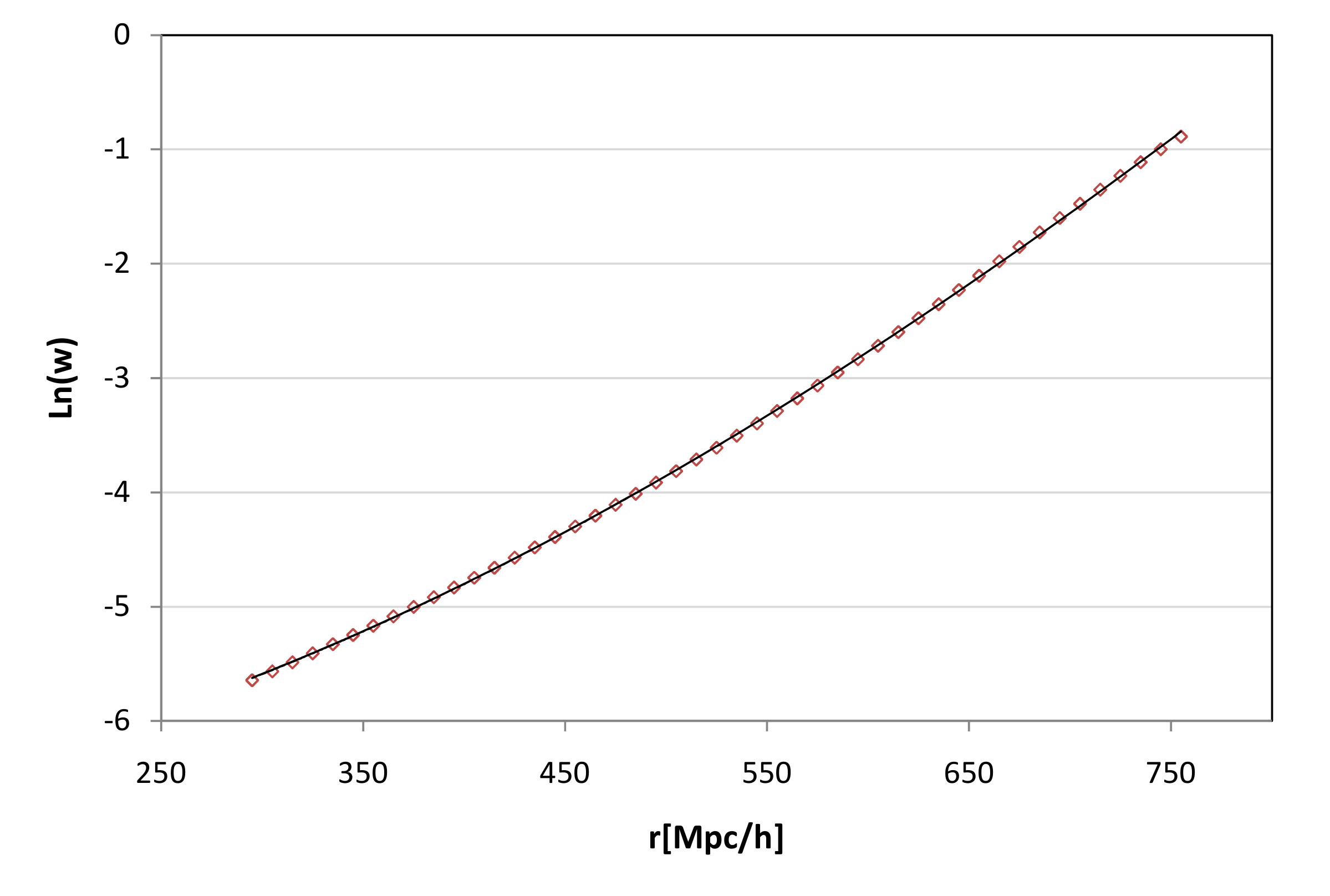}
 \end{center}
\caption{{The spatial weighting function used in our measurement,
     derived from the minimum variance estimator.}}
 \label{fig:weighting}
\end{figure}

Fig.\ \ref{fig:weighting} shows the spatial weighting function.  Due
to the large change in the weight function from the near to far edge,
we experimented with different choices for the weighting, applied to
each galaxy: (a) uniform weighting, (b) $w^{1/2}$, (c) $w$. We found
that (c) results in a substantially increased shot noise, since there
is too much weight added to the small number of objects at the far
edge of the volume, while there is very little difference between (a)
and (b).  As a result, we adopt (b), the square-root-weighted high-$z$
sample, as our fiducial one for analysis.  Indeed, cutting galaxies
closer than $300\hMpc$ does entirely remove the $170\hMpc$ feature.

\begin{figure}
  \begin{center}
    \includegraphics[scale=0.6]{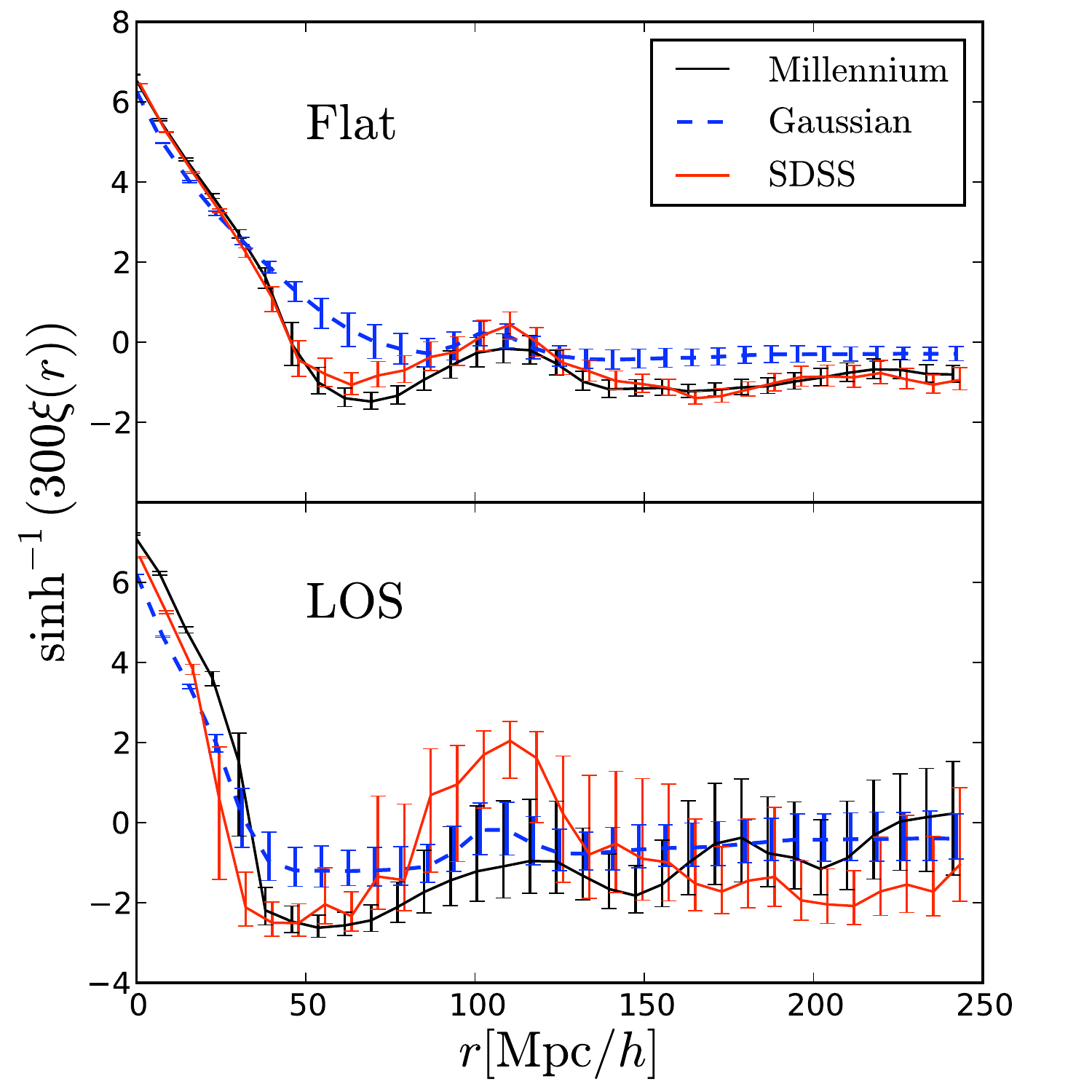}
  \end{center}
\caption{Angle-averaged correlation functions for Gaussian
    simulations, galaxies in the MS, and from the high-$z$ SDSS
    sample.  The top panel shows the $\xi$ averaged uniformly over
    angle, and the bottom uses only data within $6\degr$ of the LOS.
    The error bars are discussed in the text.}
 \label{fig:xi1d}
\end{figure}

Figure \ref{fig:xi1d} shows flat angle-averaged, and LOS, correlation
functions for MS and Gaussian samples, and for the high-$z$ SDSS
sample.  No galaxy-bias factor was applied to the Gaussian curves.

In each case, $\xi$ is measured among 2D slices, and then averaged
together.  The error bars are rather different in size; this is
because of the different sample volumes.  Also, as usual in
discussions of the correlation function, we should note that the error
bars are somewhat correlated.

The Gaussian error bars are the simulation-to-simulation standard
deviations of $\xi$, measured in 2D slices (using only one slicing
orientation) and then averaged together.  In the MS and SDSS cases,
the error bars are estimated from within the sample of slice
correlation functions, requiring an estimate of the number of degrees
of freedom (DOF) among the correlated slices.  The square root of this
DOF is the factor by which we divide the slice-by-slice dispersion to
get the plotted error bars.  

In the MS case, the flat-angular-weighting DOF is the number of slices
(128 = 2 axes $\times$ 64 slices per axis) divided by 1.2, the same
factor used below in Section \ref{sec:correlated_slices} from the
Gaussian simulations to account for the additional cosmic variance
from going from ensembles of slices to ensembles of simulations.  The
DOF in the LOS case gets divided by an additional, conservative factor
of two because many LOS pairs of galaxies are present in slices along
both slicing directions.  We were particularly careful for the SDSS
sample in estimating the reduction in DOF from correlated slicing
orientations; see Sect.\ \ref{sec:dof} below for details.

The flat-angular-weighting $\xi$'s in the SDSS and MS samples are
strikingly similar in shape.  In fact, we do not necessarily expect
the samples to match, since the galaxy properties or number densities
do not match between the MS and SDSS samples.  We consider the high
degree of agreement to be largely by chance.

One feature that seems quite solid, though, is a LOS trough at
$\pi\approx 55\hMpc$ (see also Fig.\ \ref{fig:SNraw-12}, below), that
is much deeper than the Gaussian simulations.  Multiplying the
Gaussian curve by some linear bias factor could perhaps make these
troughs line up better, but this would cause disagreement at larger
$r$, and would require quite a large bias factor, which we do not
expect for this relatively low-luminosity sample.  The depth of the
trough is a sign that non-linear infall and/or galaxy bias clear out
this region along the LOS much more dramatically than in linear
theory.  We speculate that in the SDSS case, one reason for the strong
peak along the LOS is a pile-up of these cleared-out galaxies.

\subsection{Effective degrees of freedom}
\label{sec:dof}
While our slicing strategy enables an estimate of error bars and
signal significance, this estimate is rather complicated, since slices
overlap, and even when they do not, nearby slices are likely
correlated with each other.  If all slices were statistically
independent, the degrees of freedom (DOF) would be the number of
slices, 661.  We investigate two factors by which this number must be
reduced: first, a factor coming from slice correlations and cosmic
variance; second, a factor from overlap due to the 12 different
angular slicings.

\subsubsection{Correlated slices and cosmic variance}
\label{sec:correlated_slices}
We typically expect an overestimate in the significance level of a
detection when estimating it from within the sample.  We
investigate this effect on Gaussian simulations to which linear
redshift distortions have been applied.

As in Section \ref{sec:wavgauss}, we sliced Gaussian simulations (box
size 768$\hMpc$, cell size $2\hMpc$) into 16$\hMpc$ slices.  Here, we
take just one slicing per simulation.  From each simulation, we
estimate (S/N)$_{\rm slices}$ of the wavelet coefficient at each
$(r_b,s)$ by measuring the mean and standard deviation among slices
within the simulation.  In this case, we multiply the S/N by
$\sqrt{N_{\rm slices}}$, the DOF of the single slicing in the
approximation that all slices are independent.  We also measure
(S/N)$_{\rm slicings}$ more properly, measuring the mean and standard
deviation of averages of slicings of different simulations.  The
latter estimate, which includes the effects of slice-to-slice
correlation and cosmic variance, is the same as those performed in
Section \ref{sec:wavgauss}, except that here, one (instead of two)
slicing orientation is used.

\begin{figure}
  \begin{center}
  \includegraphics[scale=0.4]{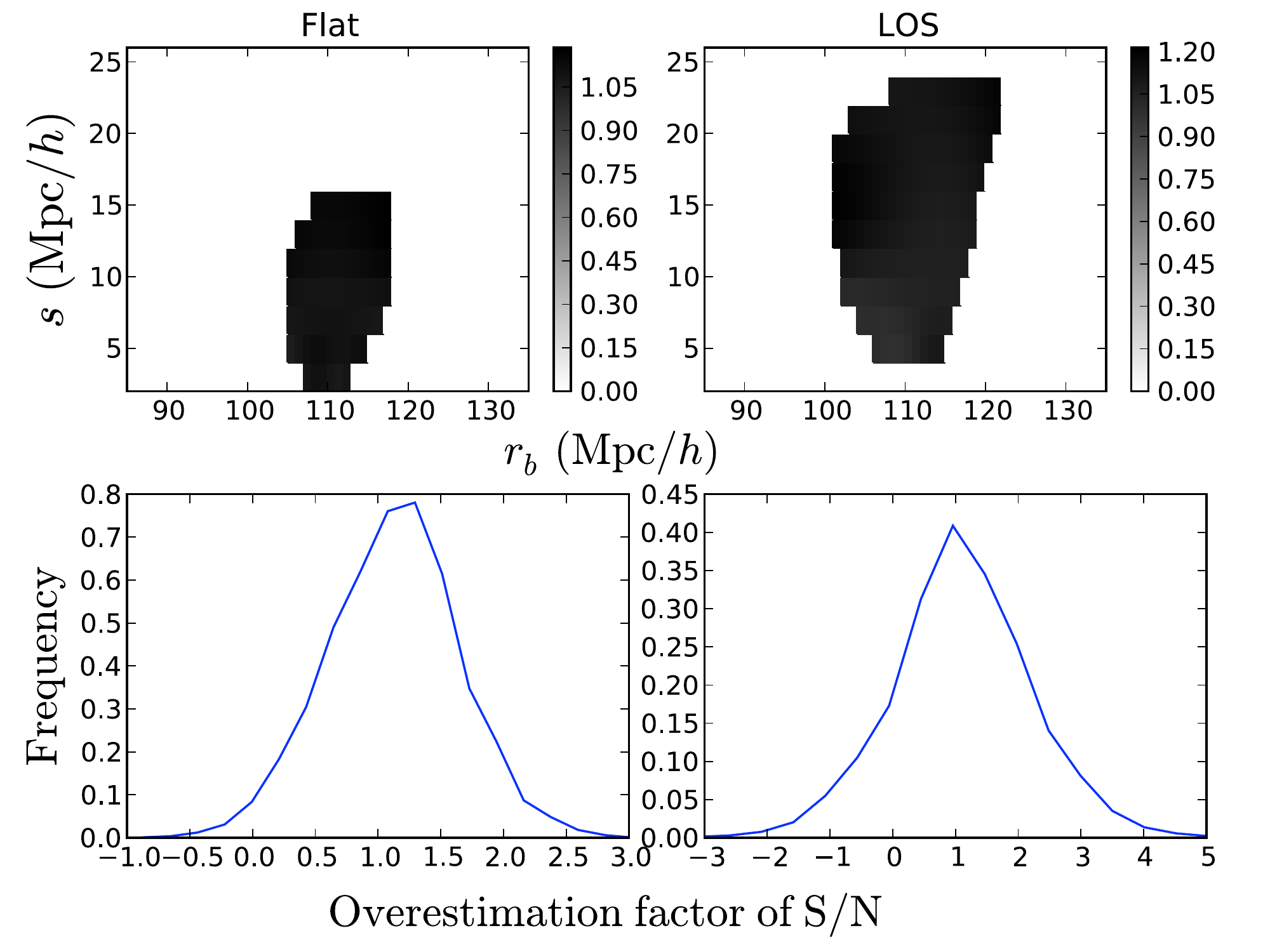}
  \end{center}
\caption{Top row: for flat and LOS weightings, the mean
    overestimation factors of the S/N levels from estimating the S/N
    within a sample of correlated slices, i.e.\ $\avg{({\rm S/N})_{\rm slices}}$/(S/N)$_{\rm slicings}$.  Outside a ``peak region''
    (with a mean S/N at least half of the maximum S/N), the plotted
    value is set to zero.  Bottom row: histograms of the
    overestimation factor, drawing from all simulations and from all
    ($r_b,s$) within the peak region.
 \label{fig:overestim}
 }
\end{figure}

Fig.\ \ref{fig:overestim} shows the means and distributions of the
ratio of (S/N)$_{\rm slices}$, which is different for each simulation,
to (S/N)$_{\rm slicings}$.  We expect it to exceed one on average,
since estimating the significance among slices should give an
overestimate.  In the top row, the dark regions are within the ``peak
region,'' defined as having a (S/N)$_{\rm slicings}$ of at least half
the maximum (S/N)$_{\rm slicings}$.  We focused on this peak region to
avoid dividing by small numbers when taking the ratio.  The bottom row
shows histograms of this ratio over all simulations, and over all
$(r_b,s)$ in the peak region.  Results are shown for both 'flat' and
LOS angular weighting.

Although we have used simulations with about the same volume as the
SDSS sample, again they are purely Gaussian simulations, with linear
redshift distortions.  The slice-to-slice variance in the wavelet
coefficient is surely underestimated using them.  However, in Gaussian
simulations, the simulation-to-simulation variance is likely
underestimated as well, so we expect their ratio (which is what
actually enters our SDSS analysis) not to be underestimated as
severely as the slice-to-slice variance in the wavelet coefficient.

The mean of this ratio of standard deviations in the peak region is
only a bit over 1; the mean in both LOS and flat weightings is 1.1.
Thus we adopt $1.1^2=1.21$ as the DOF reduction factor from going from
slice-to-slice variance to simulation-to-simulation variance.  We
should note that there is a large dispersion in this factor, which
simply means that a given simulation can have a much larger or smaller
signal than the mean would give.

\subsubsection{Correlated slicing orientations}
\label{sec:slicings}
The factor by which we would overestimate the DOF by assuming all
slicings to be independent is $\nor/\nore$, where $\nor$ is the number
of slicings, and $\nore$ is the number of slicings including
degeneracy.  Roughly, $\nor/\nore=\noc$, the mean number of slices
occupied by a pair of galaxies.  We estimate $\noc$ in two ways:
first, by counting galaxy pairs that go into the 2D and 3D correlation
functions.  Second, we relate the problem to that of Buffon's needle
\citep{buffon}.

\begin{figure}
  \begin{center}
  \includegraphics[scale=0.4]{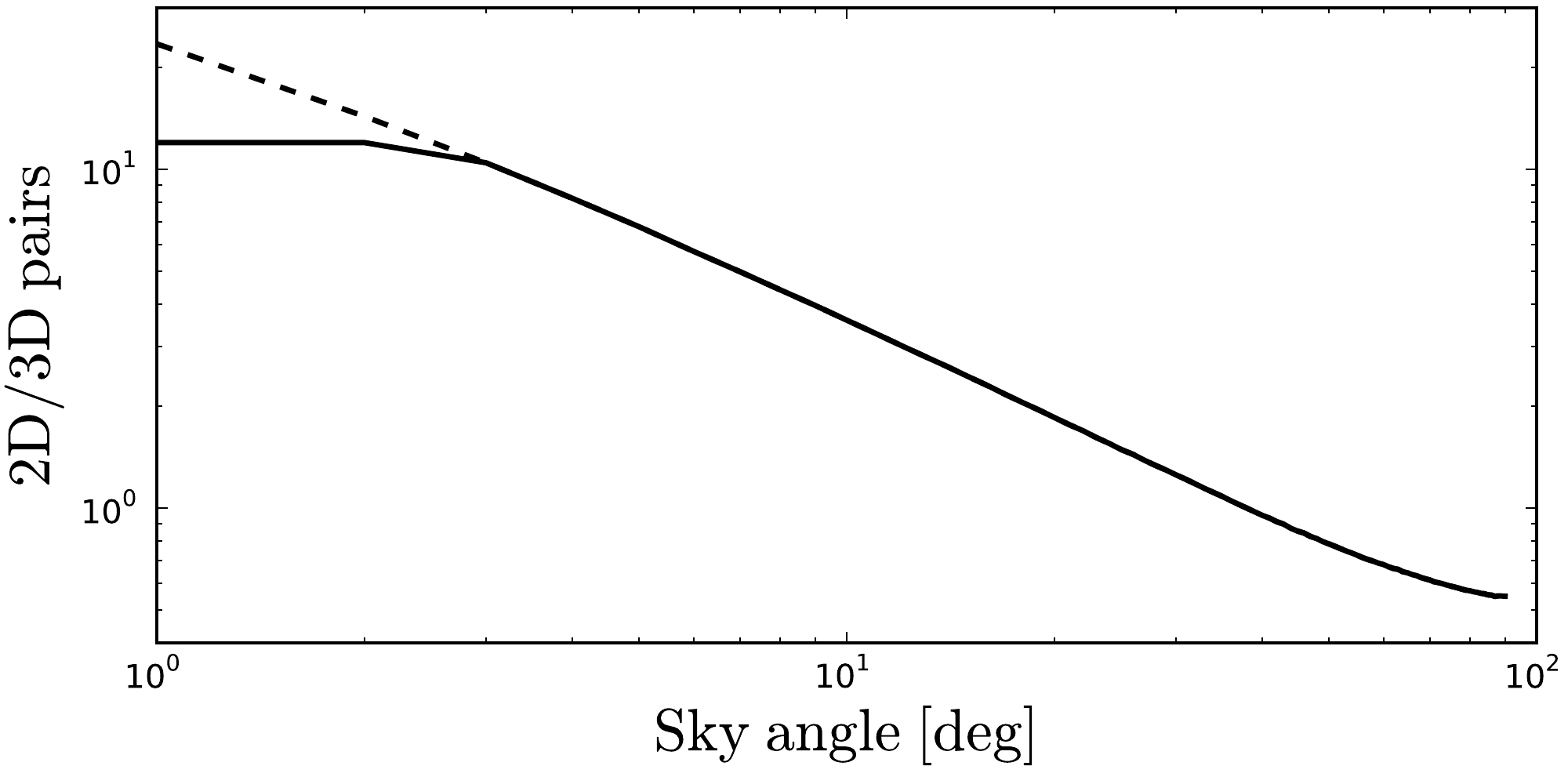}
  \end{center}
\caption{Dashed, the ratio between the number of data-data pairs
    measured in 2D (slicing and projecting), and in 3D, without any
    slicing.  Because of the projection from 3D onto the plane of the
    slice, this ratio exceeds 12 (the number of slicings) at small
    separation.  For the calculation of the DOF, we truncate this
    function at 12 (solid).
    \label{fig:ddratio}
  }
\end{figure}

To estimate $\nore$ in a brute-force fashion, we add up the raw number
of galaxy pairs ($DD$), counted in all 661 2D slices, as a function of
$(r,\theta)$ in the $(\pi,r_p)$ plane.  We also measure the pairs in
3D, and compare them.  Using a range in $r$ of 100-120$\hMpc$, we
count the pairs in 1$\degr$-wide bins.  Fig.\ \ref{fig:ddratio} shows
the ratio of the 2D to 3D pairs.  In the limit of small sky
separation, we expect all pairs to enter 12 slices.  The reason that
the plotted ratio exceeds 12 at small separation is because of the
slices' 2D projection, important for $\theta<2.5\degr$ (the slice
width): a pair's 3D $\tsky$ gets projected onto the plane of the
slice, resulting in a smaller 2D $\tsky$.  Also note that at large
$\theta$, the fraction dips below 1.  This means that the 12 slicings
were not enough to catch all pairs at large angles.

To estimate $\nore$ for the two weightings, we average the plotted
ratio from 0 to 6$\degr$ (LOS), and from 0 to 90$\degr$ (flat).  Since
pairs cannot inhabit over 12 slices, for this average we truncate the
curve at 12.  Averaging this truncated curve gives $\noc=9.2$ (LOS)
and $\noc=1.7$ (flat).

Another, analytical way of estimating the the number of slices on
average that a pair of galaxies inhabits is related to the problem of
Buffon's needle, which can be stated as follows: Let $\pbuf(\ell/t)$
denote the probability that a needle of length $\ell$, dropped
randomly on a piece of paper with parallel lines separated by a
distance $t$, will cross at least one line.  Buffon found that
\begin{equation}
  \pbuf(x=\ell/t) = \frac{2}{\pi}x,\ x\le 1.
  \label{eqn:buffon}
\end{equation}
The expression is more complicated if $x>1$, but we do not use that
case here.

Associating galaxy pairs with ``needles,'' and slice edges with the
parallel lines on the paper, a galaxy pair will inhabit a slice if it
does not cross any lines.  Imagining a slice orientation as a random
rotation and translation of the lined paper, the fraction of slice
orientations for which a galaxy pair (of separation $\tsky$ on the
sky) appears in a slice (of angular thickness $t$) will be
$1-\pbuf(\tsky/t)$.  With 12 orientations, the average number of
slices occupied per galaxy will be $\noc(\tsky)=12[1-\pbuf(\tsky/t)]$.

If the angle in the $\xi(\pi,r_p)$ plot were the angle on the sky, we
would simply average $\noc(\tsky)$ over a range of $\tsky$ to get the
average LOS galaxy-pair-slice occupancy.  We work in $(\pi,r_p)$
coordinates though, so we estimate a maximum $\tsky$ around the BAO
scale along the LOS to be about 1/3.5 of the maximum angle in
$(\pi,r_p)$.  This 1/3.5 factor is a typical ratio of their LOS
separation ($\sim 100\hMpc$) to the LOS distance to the observer
($\sim 350\hMpc$).  Averaging $12[1-\pbuf(\tsky/t)]$ over angles from
$0$ to $6/3.5$ gives $\noc=$ 9.4 slices, in accord with the previous
estimate based on pair counts.

To illustrate the degree of independence of $\xi$ among the 12 slice
orientations, we plot $\xi$ for each orientation, and also its mean.
Fig.\ \ref{fig:orient} shows this for LOS and flat weightings.  As
expected from the arguments in this Section, there is much more
variance with flat weighting than LOS, since the galaxy pairs along
the LOS occupy slices of many slice orientations.

\section{Wavelet analysis of the SDSS Results}

\begin{figure}
  \begin{center} 
    \includegraphics[scale=0.43]{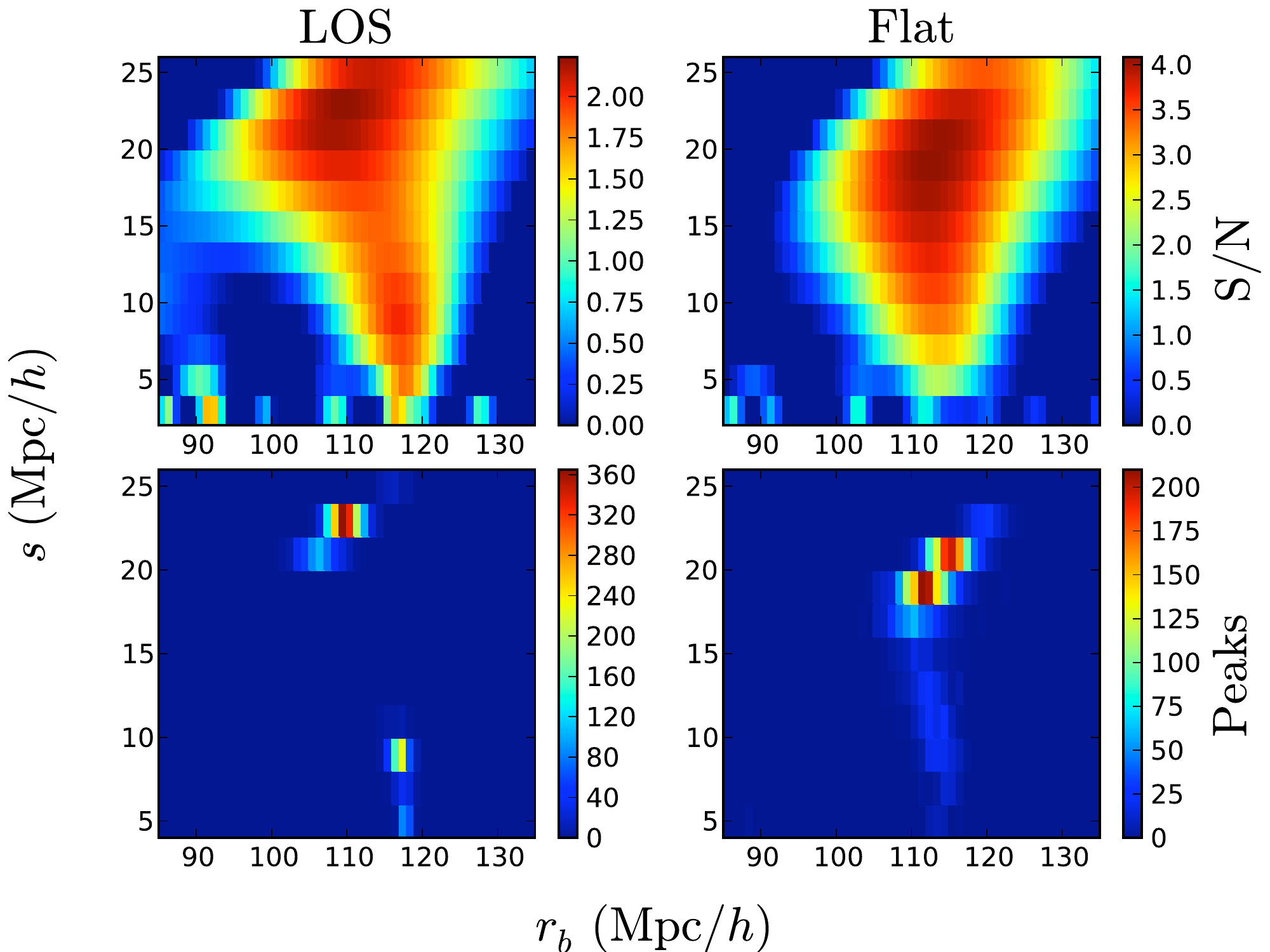}
  \end{center}
  \caption[1]{Top row: signal-to-noise plots showing the estimated
    significances of bumps in the high-redshift SDSS sample, as a
    function of $(r_b,s)$.  Bottom row: 2D histograms of the locations
    of S/N peaks, among 3000 bootstrap-resamplings.
    \label{fig:sdsspeaks}
  }
\end{figure}

We apply the same wavelet-based peak finder described in the previous
sections to the high-redshift SDSS results.  Fig.\ \ref{fig:sdsspeaks}
shows the results.  The top panels shows the S/N from this sample,
with LOS and flat angular weighting, over the same range
(i.e.\ Bayesian prior) in ($r_b,s$) as in the Gaussian simulations.
Here the raw inter-slice S/N is divided by the effective number of
degrees of freedom (DOF), discussed in Section \ref{sec:dof}.  If
the slices were independent,the DOF would simply be the number of
slices, 661.  But this needs to be reduced by a factor from slice
correlation and cosmic variance (which we estimate to be 1.2), and a
factor from physical slice overlaps from the multiple slicing
orientations (which we estimate to be 9.2 for LOS weighting, and 1.7
for flat weighting). With this correction, we estimate the peak S/N of
the LOS peak to be significant at the 2.2-$\sigma$ level, and the peak
with flat angular weighting at the 4.0-$\sigma$ level.  As in the
previous result plots, we weight each slice by its width.

The bottom panels show 2D histograms of peak locations from 3000
bootstrap-resamplings of 661 slices (with replacement) from the 661
slices.  Because we use the same number of slices in the bootstrap
samples as in the measurement, this gives an estimate of the actual
uncertainty in the peak positions, estimated from within the sample
itself.

The peak locations, with error bars thus estimated from within the
sample, are $109.7\pm 4.7$ (LOS), and $113.6\pm 3.5\hMpc$.  These
error bars consider some cosmic variance, but only that found within
the sample.  To this, we add in quadrature the error bars from cosmic
variance outside the sample, estimated from the Gaussian simulations
(which give error bars of $8\hMpc$ in LOS, and $2\hMpc$ with flat
weighting).  This gives estimates of $109.7\pm9.3\hMpc$ (LOS) and
$113.6\pm4.0\hMpc$ (flat).  These do not include systematic error
bars, which above we estimated to be about $1\hMpc$ for the Gaussian
simulations, but would likely be larger for fully non-linear
simulations.

The S/N peaks in both the MS and SDSS samples are at substantially
higher $s$ than in the Gaussian simulations.  One possible cause for
this is the broadening in the BAO feature that Lagrangian
displacements of order $10\hMpc$ from large-scale tidal motions
produce \citep[e.g.\ ][]{esw07}.  Also, fingers of God likely broaden
the peak in the LOS direction.

Of course, our Gaussian simulations we use here are idealized, missing
several complications that are mild but perhaps not entirely
negligible on BAO scales: non-linearities in matter clustering; shot
noise; effects from the non-cubic, non-periodic survey shape; bias
between the galaxy and matter fields; non-linear redshift distortions,
especially fingers of God; and perhaps large non-Gaussian (co)variance
\citep{rh05}, which however seems a bit milder for redshift-space
galaxies than for real-space matter \citep{nss10}.

To get an iron-clad estimate of the significance, we need
high-resolution Gpc-scale simulations, that would both encompass a
representative volume, and possess full non-linearities.  This will be
part of a follow-up analysis, where we will build a few hundred rather
high-resolution simulations, analyzed using realistic, SDSS-like
survey geometries.

In the mean time, because these constraints will likely enlarge in the
face of non-linearities, we conservatively adopt the looser constraint
on the BAO peak position, that comes from the LOS weighting, rounding
up the error bar to $110\pm 10\hMpc$.  This error bar is likely larger
than any additional systematic errors from non-linearities.  Also, it
would accommodate the peak in the full (including the Sloan Great Wall)
sample, which is at 97$\hMpc$.

While 4.0 is a measure of the S/N of a bump in the SDSS sample at the
particular peak ($r_b,s$) in the S/N plot, the percentages given at
the end of Section \ref{sec:freq_max} give another measure of the
significance of our peak detection, that is immune to {\it a
  posteriori} bias issues.  0.2\% of no-wiggle simulations have any
bump of S/N $\ge 4.0$ that could plausibly be associated with the BAO
feature (i.e.\ that is between 100 and 120$\hMpc$.  This implies a
99.8\% chance that a bump such as we see (i.e.\ anywhere in the
vicinity of the expected BAO location) is truly a BAO feature.

\begin{figure}
  \begin{center}
    \includegraphics[scale=0.33]{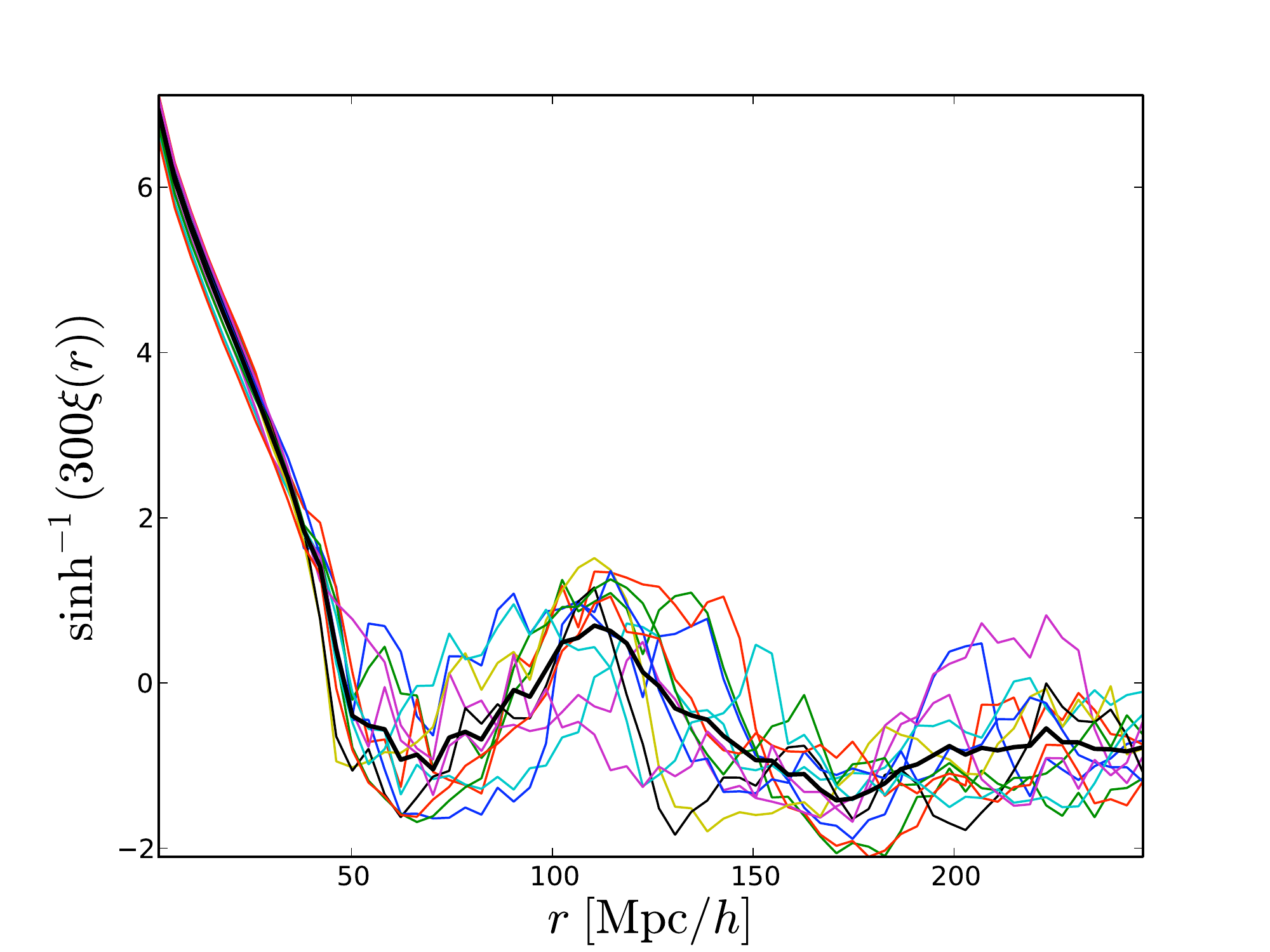}
    \includegraphics[scale=0.33]{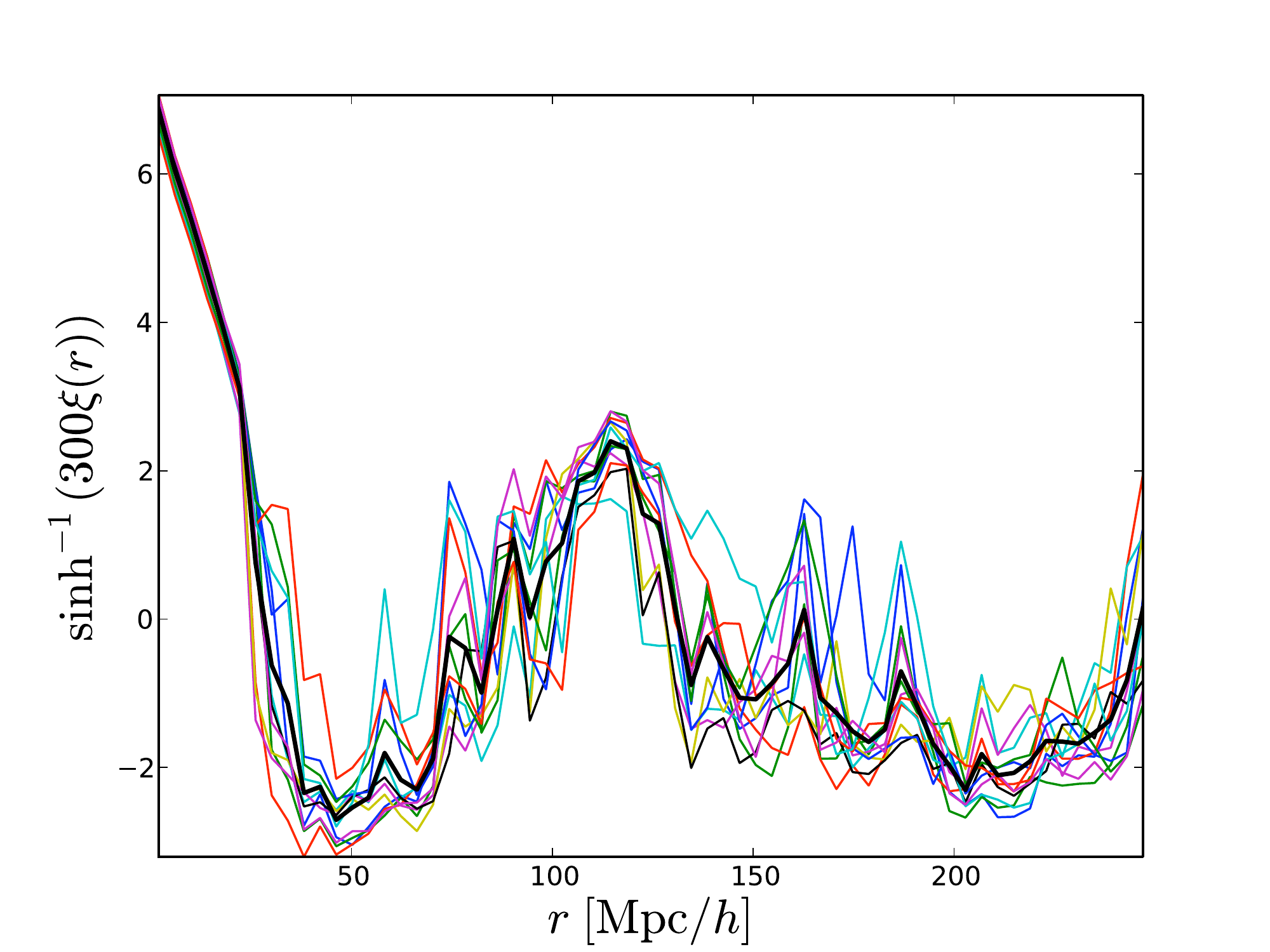}
  \end{center}
  \caption{Top: $\sinh^{-1}(300\xi(\pi,r_p))$ using flat angular
    weighting in the ($\pi,r_p$) plane, for all 12 slicing
    orientations, and their mean (black, bold).  Bottom: same, for LOS
    angular weighting.
    \label{fig:orient}
  } 
\end{figure}

In another test, that does not depend at all on Gaussian simulations,
we use the variances among the 12 slicings of the wavelet
transforms of correlation functions, averaged over all slices within
the slicing.  Fig.\ \ref{fig:orient} shows these correlation
functions, along with their mean.  As expected, the LOS dispersion is
smaller than with flat weighting, since the LOS measurements are
highly correlated from slicing to slicing.  In contrast, the predicted
simulation-to-simulation variance in the LOS is greater than with flat
weighting; this is why our stated S/N level is lower with LOS than
with flat weighting.

\begin{figure}
  \begin{center}
    \includegraphics[scale=0.45]{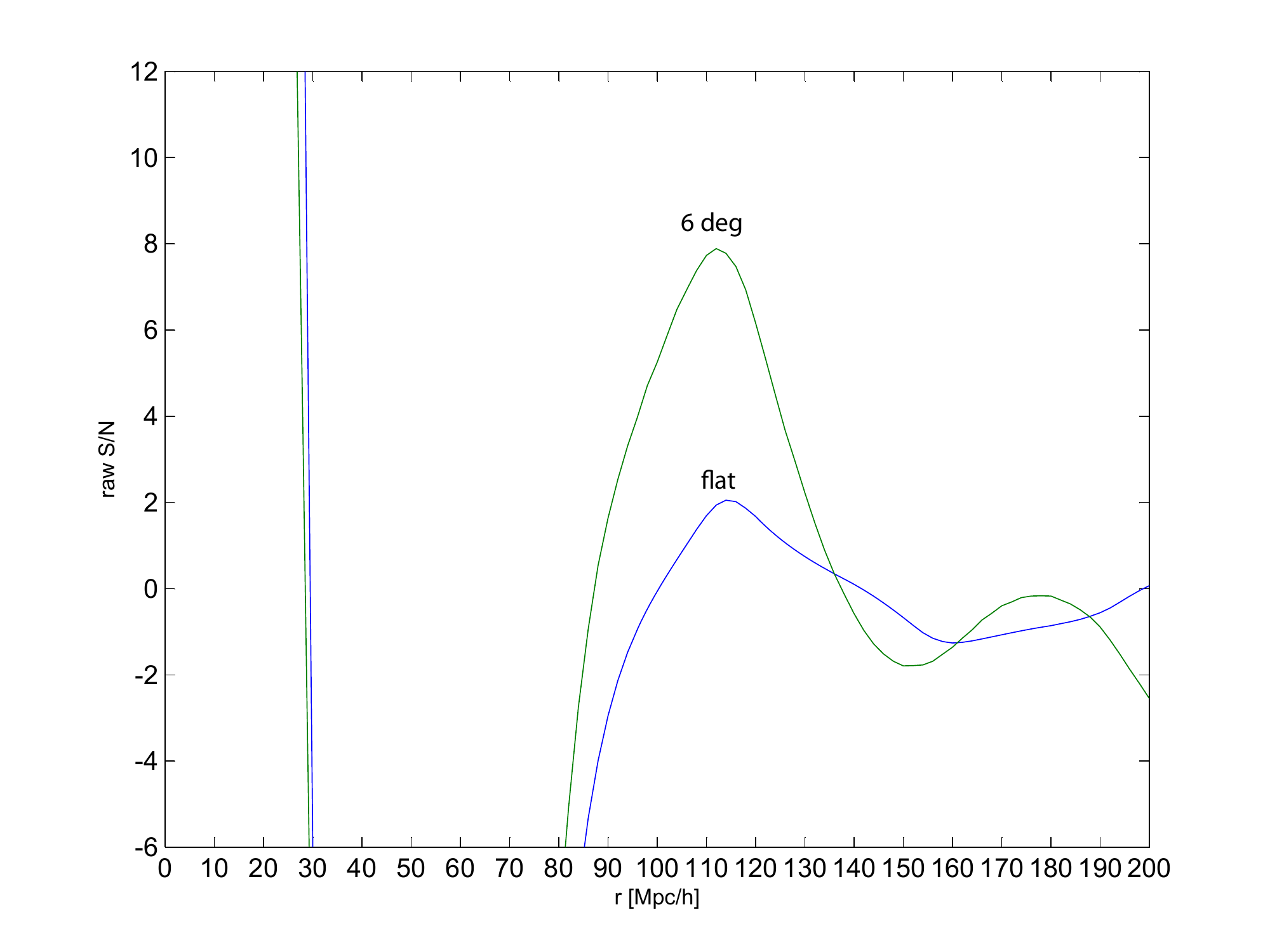}
  \end{center}
  \caption{The S/N of the wavelet transform as a function of wavelet
    peak location $r_b$, using the slicing-to-slicing dispersion for
    the noise N.  The wavelet scale is fixed at $s=20\hMpc$, where the
    peak over all ($r_b,s$) is found.
    \label{fig:SNraw-12}
  } 
\end{figure}

Fig.\ \ref{fig:SNraw-12} shows the wavelet transform's S/N as a
function of $r_b$, holding the wavelet scale fixed at $s=20\hMpc$,
using the raw, slicing-to-slicing dispersion for the noise N.  This
gives a LOS peak S/N of 7.9, and a flat-weighting S/N of 1.9.  This is
without reducing the noise (by the square root of the effective DOF)
because we are averaging the results of different slicings together.
The effective DOF is the effective number of independent slicings;
These are $12/9.2=1.3$ (LOS), and $12/1.7=7$ (flat), bringing the
estimated S/N to 9.0 (LOS) and 5.2 (flat).  The LOS significance, in
particular, seems ridiculously large, but we must remember that it is
based on only 1.3 effective measurements; including the error bar
(fractionally, $1/\sqrt{{\rm DOF}}$), the S/N is $9.0\pm 7.9$.  The
slicing-to-slicing estimate of the flat S/N is more meaningful,
$5.2\pm 2$.  We should note that this estimate does neglect
slicing-to-slicing correlation that goes beyond actual overlaps of
galaxy pairs, but we expect this additional correlation to be small.

Fig.\ \ref{fig:SNraw-12} also shows the tremendous significance of the
trough at $r_b\approx 55\hMpc$.  We speculate that this trough is
cleared out to a greater degree in the presence of a BAO peak, and
this is an interesting issue for further study.

\begin{figure}
  \begin{center}
    \includegraphics[scale=0.45]{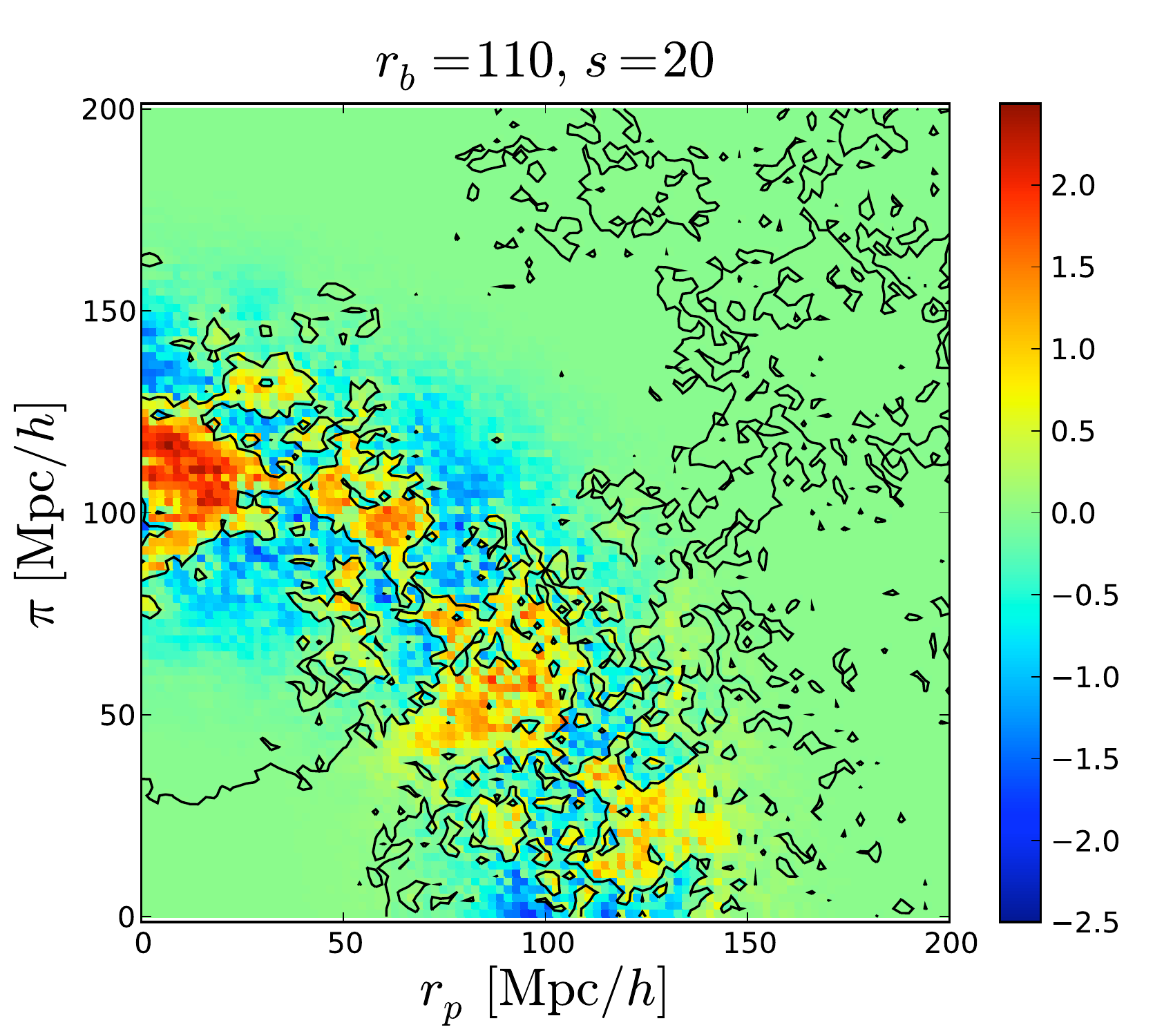}
  \end{center}
  \caption{$\xisdss(\pi,r_p)$ flattened, and attenuated with a radial
    Gaussian window, at $(r_b=110,s=20)$.  This visual transformation
    emphasizes the information to which the Mexican Hat wavelet is
    sensitive.
    \label{fig:sdss_sqrtw_rb110_s20}
 }
\end{figure}

Fig.\ \ref{fig:sdss_sqrtw_rb110_s20} shows $\xi$ from the SDSS sample,
flattened in the manner of Sec.\ \ref{sec:flattening}, with parameters
at the flat-weighting peak ($r_b,s$) in Fig.\ \ref{fig:sdsspeaks}.

\section{Conclusion}

In this paper we have shown how redshift space distortions can amplify
and modify features in the galaxy correlation function.  In linear
theory, the BAO feature sharpens as the angle approaches the line of
sight.  While we expect nonlinear effects such as fingers of God to
broaden the BAO feature along the line of sight (LOS), they do not
seem to dampen it substantially.

We see a compelling LOS ``bump'' in the SDSS DR7 MGS correlation
function, measured in 2D slices of varying pitch angle that cover the
sample.  We analyze the bump using a Mexican-hat wavelet transform,
which has a peak in signal-to-noise ratio S/N at $(110\pm 10)\hMpc$,
marginalized over the scale of the wavelet.  We assess the
significance of the bump by looking at variances of slice correlation
functions within the sample, which presents difficulties because of
slice-to-slice correlations.  Nevertheless, at the particular location
and wavelet scale of the peak in wavelet S/N, we tentatively assess
the significance of this peak at 4$\sigma$ if the correlation function
is averaged in a flat angular weighting, and at 2.2$\sigma$ if it is
averaged only within $6\degr$ of the line of sight.

These significance estimates employ Gaussian simulations, and thus are
likely a bit optimistic.  However, this optimism is more modest than
it might at first seem; we only use the Gaussian simulations to
extrapolate the significance estimate measured from slices within the
sample to the proper, sample-to-sample significance estimate.  We make
another estimate that does not use simulations at all.  From just the
variance of the wavelet transform among angular slicing orientations,
we estimate that the S/N $=9\pm 8$ (LOS) and S/N $=5\pm 2$ (flat
weighting).

Taking into account {\it a posteriori} bias, we also ask how common a
bump of the measured significance is in Gaussian no-wiggle
simulations, not just at the peak of the wavelet S/N, but anywhere in
the vicinity of the BAO feature (within 10$\hMpc$ of 110$\hMpc$).
Evaluating the wavelet using flat angular weighting, we find that such
a bump is only present in 0.2\% of no-wiggle simulations.

It is curious that the LOS BAO feature seems to be stronger than
apparently expected from simulations, in both the Sloan LRG
\citep{g09} and main-galaxy samples.  Its presence in both, rather
independent, samples suggests that this is not a fluke.

We propose that the $\sin\theta$ weighting that is commonly used,
while sensibly motivated since it follows the distribution of galaxy
pairs in a 3D sample, is likely suboptimal in constraining power,
since it greatly suppresses any LOS signal.  An example of an
alternative weighting that is less hostile to the LOS is the flat
weighting we employ, which is natural for the 2D slicing strategy we
use.

While our current analysis makes the case that the features we see are
rather statistically significant, a conclusive determination awaits a
planned study employing more-realistic simulations, that include
nonlinearities, galaxy bias, survey-shape effects, etc.  This study
will not only address our present result specifically, but we hope
will provide a definitive answer the question of how BAO constraining
power varies with angle from the LOS.

Our claimed rather high significance level may be surprising in light
of recent analyses casting doubt on the reality of BAO detections even
in the larger, SDSS LRG sample \citep{kazin10,cabre10}.  In these
analyses, if a correlation function is better-fit by a ``no-wiggle''
model, there is deemed not to be a bump.  Although we do not
quantitatively compare the methods, our peak detector is likely more
tolerant of modest bumps.  For us, essentially there exists a bump at
a certain wavelet scale and location if the correlation function has a
positive wavelet coefficient there, i.e.\ that the second derivative
of the correlation function, smoothed over the wavelet scale, is
negative.

Even if we have overestimated the significance of our detection, an
important point is that it still may be used for cosmological
constraints.  As \citet{cabre10} emphasize, the hypothesis test that a
BAO feature exists in a given sample is a separate statistical
question from the degree of its cosmology-constraining power.  There
is essentially no doubt that the power spectrum underlying the
structure in our Universe had a BAO feature at the epoch when the
cosmic microwave background was emitted \citep[e.g.][]{larson10}, and
if it is not present at the present epoch, that would be a big
cosmological puzzle.

Under the assumption that the power spectrum underlying the structure
in our Universe has a BAO feature, we find that the appearance of even
a low-significance bump gives some constraining power over the BAO
feature's location.  For example, in a Gaussian sample with the volume
of ours, if a clear bump exists in a sample (which happens about half
the time), the bump gives an error bar in the BAO peak location of
$\sim8\hMpc$.

\acknowledgments

We thank S\'{e}bastien Heinis and Eyal Kazin for helpful discussions,
and the anonymous referee for a thorough and helpful report.  HJT
thanks Y.H.\ Zhao, X.L.\ Chen, and X.P.\ Zheng for support and
guidance.  We are also grateful for funding from the W.M. Keck and
Gordon and Betty Moore Foundations.  The Millennium Simulation
databases used in this paper and the web application providing online
access to them were constructed as part of the activities of the
German Astrophysical Virtual Observatory.

Funding for the SDSS and SDSS-II has been provided by the Alfred
P. Sloan Foundation, the Participating Institutions, the National
Science Foundation, the U.S. Department of Energy, the National
Aeronautics and Space Administration, the Japanese Monbukagakusho, the
Max Planck Society, and the Higher Education Funding Council for
England. The SDSS Web Site is http://www.sdss.org/.

The SDSS is managed by the Astrophysical Research Consortium for the
Participating Institutions. The Participating Institutions are the
American Museum of Natural History, Astrophysical Institute Potsdam,
University of Basel, University of Cambridge, Case Western Reserve
University, University of Chicago, Drexel University, Fermilab, the
Institute for Advanced Study, the Japan Participation Group, Johns
Hopkins University, the Joint Institute for Nuclear Astrophysics, the
Kavli Institute for Particle Astrophysics and Cosmology, the Korean
Scientist Group, the Chinese Academy of Sciences (LAMOST), Los Alamos
National Laboratory, the Max-Planck-Institute for Astronomy (MPIA),
the Max-Planck-Institute for Astrophysics (MPA), New Mexico State
University, Ohio State University, University of Pittsburgh,
University of Portsmouth, Princeton University, the United States
Naval Observatory, and the University of Washington.

\bibliographystyle{hapj}
\bibliography{refs}

\end{document}